\documentclass[twocolumn,aps,prx,longbibliography]{revtex4-2}

\usepackage{graphicx,color,hyperref}

\usepackage{amsmath}
\usepackage{amsfonts}
\usepackage{amssymb}
\usepackage{dsfont}

\newcommand{\tr}{\mathrm{tr}}

\newcommand{\bs}{\boldsymbol}

\newcommand{\beq}{\begin{eqnarray}}
\newcommand{\eeq}{\end{eqnarray}}
\newcommand{\dg}{\dagger}
\newcommand{\bpm}{\begin{pmatrix}}
	\newcommand{\epm}{\end{pmatrix}}

\newcommand{\ket}[1]{| #1 \rangle}
\newcommand{\bra}[1]{\langle #1 |} 
\newcommand{\dirac}[2]{\langle #1 | #2 \rangle}

\begin{document}

\title{
Floquet simulators for topological surface states 
in isolation
}

\author{Kun Woo Kim$^{1,2}$, Dmitry Bagrets$^1$, Tobias Micklitz$^3$, Alexander Altland$^1$
}
\affiliation{
$^1$Institut f\"ur Theoretische Physik, Universit\"at zu K\"oln, 
Z\"ulpicher Stra\ss e 77, 50937 K\"oln, Germany
\\
$^2$ Department of Physics, Chung-Ang University, 06974 Seoul, Republic of Korea \\
$^3$Centro Brasileiro de Pesquisas F\'isicas, Rua Xavier Sigaud 150, 22290-180, Rio de Janeiro, Brazil
}

\begin{abstract}

We propose dynamical protocols allowing
for the engineered realization of  topological surface states in isolation.
Our approach builds on the concept of synthetic dimensions 
generated by driving systems with incommensurate
frequencies. As a concrete example, we consider $3d$ topological surface
states of a $4d$ quantum Hall insulator via a $(1+2_\mathrm{syn})$-dimensional
protocol. 
We present first principle analytical
calculations 
demonstrating that no supporting $4d$ bulk phase is required
for a $3d$ topological surface phase. 
We back the analytical 
approach by numerical simulations and present a
detailed blueprint for the realization of the synthetic surface phase  
with existing
quantum linear optical network device technology. We then discuss generalizations, including 
a proposal for  a quantum simulator of 
the $(1+1_\mathrm{syn})$-dimensional surface of the common $3d$ topological insulator.

\end{abstract}

\date{\today}

\maketitle

\section{Introduction} 
\label{section_introduction}

Surface states of topological insulators (TI)
define one of the most  fascinating forms of quantum  matter. Depending on their symmetries and
dimensionality, they conduct charge, spin, or heat with topological protection
against the detrimental effects of impurity scattering or interactions. 
These features make the TI surface distinct from any other form of quantum matter, 
and are believed to  harbor far-reaching potential future 
device applications. At the same time, our understanding of the TI surface physics 
remains incomplete, both experimentally and theoretically. For example, 
even in the absence of interactions their conduction properties are not known quantitatively, 
and according to recent
numerical work~\cite{Sbierski:2020} even enigmatic. The experimental analysis of surface transport
is hindered by the inevitable presence of an ``insulating'' bulk, with
quotation marks because  heat or electric currents easily leak away from
the surface hindering a clear separation of surface and bulk currents.

According to the bulk--boundary principle, no \emph{lattice} quantum system in isolation 
can be in
the universality class of the TI surface. The necessity of a supporting bulk follows
from topological band theory or,  more fundamentally, as a consequence of anomaly
inflow. The main message of this paper is that this  no-go theorem can be sidestepped
within the wider framework of Floquet quantum matter. Specifically, we will propose
realizations of (dynamical) synthetic matter in universality classes
indistinguishable from those of \emph{isolated} (static) TI surfaces in the presence
of effective disorder. Our work includes three novel conceptual elements: (i) the
encoding of two and three dimensional surface state topologies in multi-frequency
dynamical protocols, (ii) the first principle demonstration of the equivalence
between the quantum states engineered in this way and insulator surface states, and
(iii) the formulation of a detailed experimental blueprint suggesting that this
program can be implemented in realistic devices within the framework of current date
technology.

 Previous work~\cite{sun2018three,higashikawa2019floquet} indeed pointed out  the realizability of  topological metallic phases in 
 dynamically driven lattice systems. However, the presence of a lattice structure made  these systems subject to the notorious fermion doubling principle, which requires  an even number of Dirac cones in the Floquet Brillouin zone. 
 In the presence of impurities these mutually gap out, spoiling the surface state analogy. 
In order to realize a genuine surface state in isolation, a more radical departure
from the solid state crystal paradigm is required. In this paper, we demonstrate
that the toolbox of quantum optics contains platforms that are up to this task,
optical lattices~\cite{chabe2008experimental, lemarie2009observation}, or linear optical networks~\cite{schreiber20122d, lorz2019photonic,geraldi2021transient} driven by multiple incommensurate frequencies. 
The driving of $d$-dimensional realizations of such systems
by  $d_\mathrm{syn}$
incommensurate frequencies is microscopically identical to a time periodic (Floquet) 
dynamics acting in an effective system of dimensionality 
$d+d_\mathrm{syn}$~\cite{Casati1989}, where the structure of the Floquet operator in the
$d$ physical and $d_\mathrm{syn}$ synthetic dimensions depends on 
the driving protocol. Importantly, the correlations in the synthetic directions are
not confined by the  fermion doubling theorem, and this will be key to the
engineering of topological surface states in isolation.  We will label the $d+d_\mathrm{syn}$-dimensional 
Floquet metallic (FM) systems realized in this way as FM$_{d+d_\mathrm{syn}}$ throughout.

The  simulation of higher dimensional systems via driven low dimensional physical
platforms is experimental reality. In breakthrough experiments it was applied to  extend one-dimensional Anderson localization in the quantum kicked rotor~\cite{Haake,shepelyansky1986localization,moore1994observation,tian2010theory} to higher dimensions. This defined an 
effectively disordered FM$_{1+2}$ and led to the first high precision observation of a three-dimensional Anderson transition under parametrically controlled conditions~\cite{chabe2008experimental,Lemarie2010}. 

However, the realization of the TI surface states addressed in this paper requires
the additional structure of an internal bi-valued degree of freedom or 'spin'. (For earlier proposals to realize topological quantum matter with synthetic dimensions via the driving of systems with internal degrees of freedom, see Refs.~\cite{Tian:2016} and \cite{martin2017topological}. Specifically,  we need full control over the lattice nearest neighbor hopping
for a system with two internal degrees of freedom (``spin''). The required technology
is not yet realized for the optical lattice~\cite{meier2018observation} but is available in the alternative
platform of linear optical networks~\cite{schreiber2010photons}. We will therefore focus on this hardware and
discuss the implementation of a FM$_{1+1_{\rm syn}}$ and a FM$_{1+2_{\rm syn}}$ TI surface state. We will
demonstrate by numerical control simulations that unique signatures of surface state
delocalization in the synthetic disordered system are observable for experimentally
accessible time and length scales. 

A further  hallmark of our approach is that it realizes surfaces in effectively 
``disordered'' phases lacking translational invariance. The reason is that the generation of synthetic dimension requires
non-commuting operators  both in synthetic and physical space. The simultaneous presence of these operators in the dynamics leads to
non-integrability and chaotic fluctuations, physically equivalent to tunable disorder
at mesoscopic length scales. Our approach, thus, simulates the surfaces of disordered 
topological quantum matter, which one may take as an added element of realism.

The plan of the paper is as follows. In Section~\ref{dynamical_protocols}, 
we present dynamical protocols of 1$d$ quantum walks that utilize synthetic dimensions to simulate higher dimensional systems. 
In Section~\ref{subsection_fm1+2} we introduce a quantum 
simulator of topological insulator surface states in isolation at the example of the 
$4d$ quantum Hall insulator. 
We introduce a $1d$ quantum walk protocol, 
discuss its topological property, and report on 
numerical simulations of the protocol, all supporting the idea 
that the surface states 
of the $4d$ quantum Hall insulator can be simulated by the $1d$ quantum walk. 
We then discuss a concrete blueprint, realizing the quantum walk within existing 
 optical linear network set-up. 
Section~\ref{subsection_simulator_quantum_spin_hall} 
provides  further details and 
discusses generalizations.
Specifically, we introduce a simulator of surface states of a $3d$ quantum spin Hall 
insulator.
We conclude in Section~\ref{section_discussion}. 
Aiming to keep the presentation as non-technical as possible, 
the details of various derivations are relegated to the Appendices.

\begin{figure}
\centering
\includegraphics[width=8.8cm]{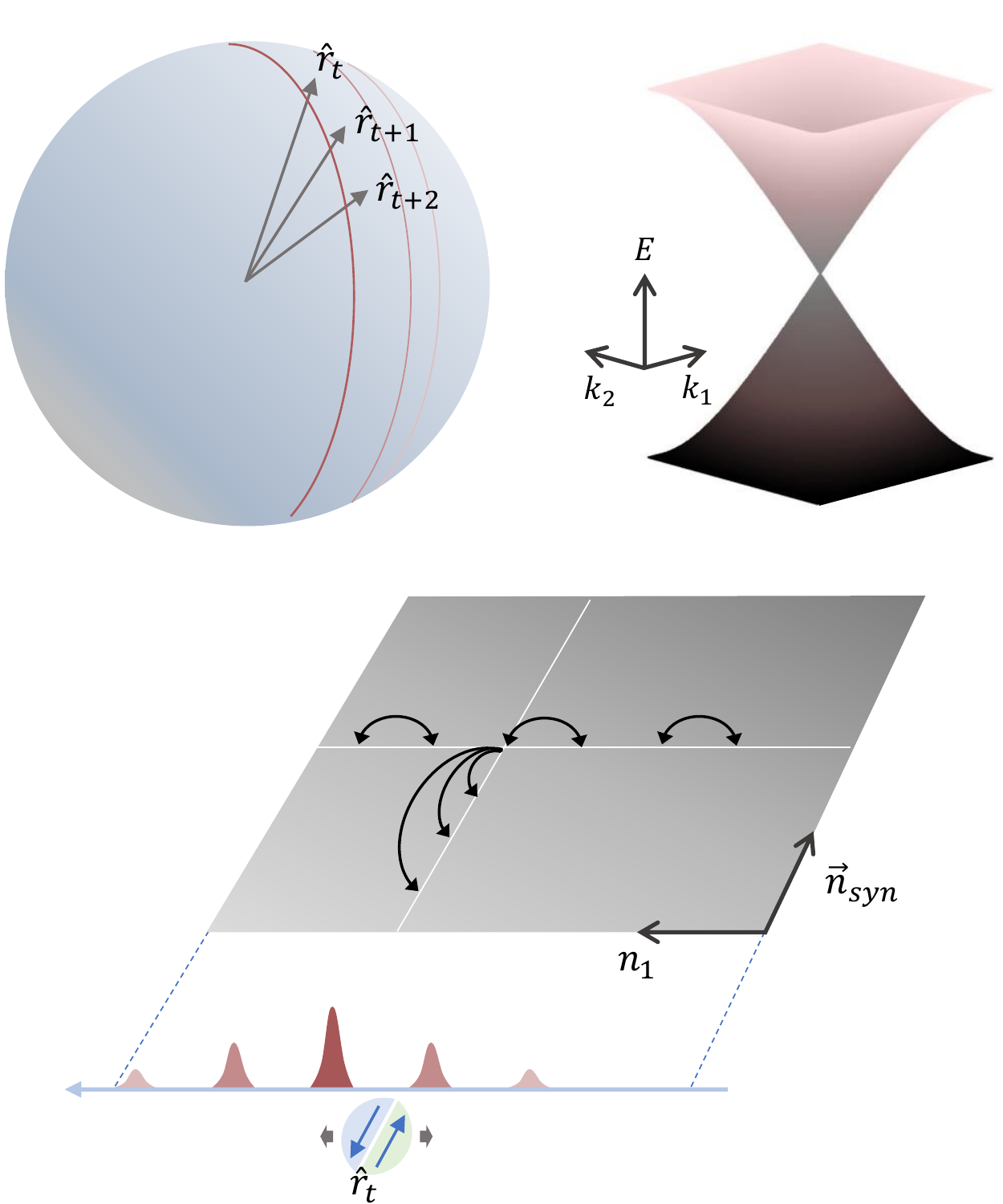}
\vspace{-0.5cm}
\caption{\label{fig:chiral_quantum_walk} 
A $1d$ quantum walk with time-dependent spin quantization axis $\hat r_t$
can simulate dynamics of a  $2d$ systems when the 
period of $r_t$  
is incommensurate with the discrete time-steps of the evolution operator. 
This engineering of synthetic dimensions 
allows to sidestep the fermion doubling principle, and to simulate 
e.g. the $2d$ surface states in isolation of a $3d$ quantum spin Hall insulator.  
}
\end{figure}

\section{Dynamical protocols}
\label{dynamical_protocols}

Consider the quantum walk of a spin-$1/2$ particle on a $1d$ lattice, generated
 by successive applications of translations and spin rotations.
 The single time-step evolution operator 
 is of the general form
\begin{align}
\hat U_t = \sum_m \hat R_m(t) \otimes \hat T_m, \label{U_step_generalization}
\end{align}
where $\hat T_m$ shifts the walker by $m$ lattice sites, and $\hat R_m(t)=\vec r_m (t) \cdot \vec \sigma$ rotates its spin. Here and in the following $\vec \sigma=(\sigma_0,i\bs\sigma)$ and $\vec r_m=(r_{m0},\bs r_m)$ are four component vectors such that 
$\sigma_0 = \mathds{1}_{2 }$, 
$\bs\sigma = (\sigma_x, \sigma_y, \sigma_z)$ and
 $\vec r_m \in \mathds{C}^4$.
Central for our proposal is the  
time-dependent  spin rotation axes $r_m(t)$ which are dynamically changed in the course of the walk.  As we show below, 
using dynamical protocols with periods that are incommensurate with the 
 discrete time step of the evolution operator Eq.~\eqref{U_step_generalization}
  enables the simulation of dynamics in higher dimensional systems.
  In the following we will focus on 
 quantum walks with short-range hops to the nearest neighbors, $m\in \{0,-1,+1\}$. The unitary operator Eq.~\eqref{U_step_generalization} then simplifies to
\begin{align}
\label{U_step_reduced}
\hat U_t=  \vec r_0 \cdot \vec \sigma  +  (\vec r_+ \cdot \vec \sigma) \otimes \hat T_+ + (\vec r_- \cdot \vec \sigma)  \otimes \hat T_- , 
\end{align}
and unitarity sets the following
constraint on $\vec r_0$, $\vec r_\pm$:
Expressing
$\vec r_\pm = \vec r_r \pm i \vec r_i$, with $\vec r_{r,i} \in \mathds{R}^4$ 
real four component vectors,
the latter are orthogonal, $\vec r_r \cdot \vec r_i =0$, $\vec r_0 \cdot \vec r_{r,i}=0$, and equal in magnitude, $|\vec r_r|=|\vec r_i|=\frac{1}{2}\sqrt{1-|\vec r_0|^2}$ at each time step $t$ (see~Appendix~\ref{quantum_walk_operator} for details).

Finally, we add spin-dependent spatial disorder to the dynamics. To this end we 
introduce the unitary matrix  
$(\hat U_{\text{dis}})_{nn'}= \hat U_{\mathrm{dis}}(n) \delta_{nn'}$
where $\hat U_{\mathrm{dis}}(n)$ are 
independent random 
spin rotation matrices, acting locally on each site $n$. 
The single time-step evolution 
$\ket {\psi_t} =  \mathcal U_{t,t-1} \ket {\psi_{t-1}}$ 
generating the $1d$ quantum walk 
is then composed of the combined 
operator, 
\begin{align}
\label{eq:CalUDef}
{ \mathcal{U}}_{t,t-1} = \hat U_t \hat U_{\text{dis}},
\end{align}
and we next discuss its potential to simulate higher dimensional dynamics.

\subsection{Synthetic dimensions from multi-frequency dynamical protocols}
\label{subsection_synthetic_dimensions}

Let us specify the protocol Eq.~\eqref{U_step_generalization} 
to rotations $\hat R_m$ which depend on $d_{\rm syn}$ time-dependent functions,
\begin{align}
\label{f}
\hat R_m(t)
\equiv \hat R_m(\varphi_{2,t},\ldots \varphi_{{d_{\rm syn}}+1,t}),
\end{align}
and where the time-dependence for each of the functions is of the form $\varphi_{i,t}=k_{i}+\omega_i t$  with frequencies $\omega_2,...,\omega_{d_{\rm syn}+1}$ incommensurate  to $2\pi$ and among themselves. 
Further, $k_\mathrm{syn}\equiv (k_{2},..,k_{d_{\rm syn}+1})$ are arbitrary initial phases which we consider averaged over in our dynamical protocols below. The
mapping to an effectively $1+d_{\rm syn}$ dimensional Floquet system is achieved by
extending the Hilbert space of the system and
interpreting these phases as momenta conjugate to integer valued coordinates $n_\mathrm{syn}=(n_2,...,n_{d_{\rm syn}+1})$, with canonical commutation relations $[\hat n_i,\hat k_j]=-i\delta_{ij}$ between the corresponding operators.
We note that $\hat n_i=-i\partial_{k_i}$ in the phase-momentum representation of the theory.   These coordinates extend the lattice in $1+{d_{\rm syn}}$ dimensions
with sites $\textbf{n}=(n_1,n_\mathrm{syn})$, where $n_1=n$ is the physical lattice coordinate,
conjugate to a phase $k_1$. In the same notation, $\textbf{k}=(k_1,k_{\rm syn})$.

Using the general relation $e^{ia \hat n}f(\hat k) e^{-ia \hat n}=f(\hat k+a) $, the time dependence in the arguments of the rotation operator can be removed by the  gauge transformation
\begin{align}
\label{eq:R_m}
\hat R_m (t) = 
e^{i\omega_j t \hat n_j} R_m(0)  e^{-i \omega_j t \hat n_j},
\end{align}
where a summation over  $j=2,\ldots,1+d_{\rm syn}$ is implicit. 
This enables us to express the time evolution operator $\mathcal U_{t,0}\equiv \mathcal U_{t,0}\equiv \prod_{\tau=0}^{t-1} \mathcal U_{\tau+1,\tau}$ as
\begin{align}
\label{eq:psi_t}
\ket {\psi_t} &= \mathcal U_{t,t-1}\mathcal U_{t-1,t-2} \cdots \mathcal U_{1,0} \ket {\psi_0}\nonumber\\
&= e^{i\omega_jt \hat n_j}\left[\mathcal U_{0,-1} e^{-i \omega_j \hat n_j }\right]^t \ket {\psi_0}.
\end{align}
We notice that the time evolution is governed by powers of the single \textit{Floquet operator} $\mathcal U_F\equiv \mathcal{U}_{0,-1} e^{-i \omega_j \hat n_j }=\hat U_{t=0} \hat U_\mathrm{dis}e^{-i \omega_j \hat n_j }\equiv U_\textbf{k}W_\textbf{n}$. Here, $\hat W_\textbf{n}\equiv \hat U_\mathrm{dis}(n_1)e^{-i \omega_j \hat n_j }$ is diagonal in the coordinate representation, while $\hat U_\textbf{k}= \hat U_0(\textbf{k})$ is momentum-diagonal. To understand this last statement, we note that in Eq.~\eqref{U_step_generalization} the coordinate translation operator $\hat T_m f(n_1)=f(n_1 - m)$ affords the representation $\hat T_m=e^{ i m \hat k_1}$ while $\hat R_m(0)$ depends on the phases $k_\mathrm{syn}$.

To summarize, our dynamics is governed by the effective multi-dimensional Floquet operator  $
\mathcal{U}_F =  \hat U_\bold{k} \hat W_\textbf{n}$ 
factoring into two pieces which are individually diagonal in coordinates and momenta, respectively. 
 Our  numerical
simulations below demonstrate  that the combined action of these 
operators induces integrability breaking, physically equivalent to 
 static disorder, in all $1+d_{\rm syn}$ dimensions. 
However, before  introducing quantum simulators  for the combined effects of disorder and topology in this setting,  
 we briefly introduce observables
 probing topological surface states in an experimentally accessible way.

\subsection{Observable} 
\label{subsection_observable}

The spreading 
 after $t$ time steps 
of a wave packet, describing a quantum walker initially prepared at site $n_1=0$ with spin $\sigma$, 
can be expressed as 
\begin{align}
    \label{cf}
     \langle \Delta X^2 \rangle
    &\equiv 
    \sum_{n_1} \sum_{\sigma',\sigma} n_1^2
   \overline{ |\langle n_1,\sigma' | \mathcal{U}_{t,0} |0,\sigma\rangle|^2 }.
\end{align}
Here the  sum is
  over  spin orientations $\sigma=\uparrow,\downarrow$ and
 $\overline{(...)}$  refers to the average over both, an ensemble of realizations of the  random rotations $\hat U_{\rm dis}$ and 
 the initial momenta $k_\mathrm{syn}$.
  In a mixed coordinate-momentum representation, basis states of the extended Hilbert space are defined  by the kets  
$|n_1,\sigma\rangle \to |n_1,k_\mathrm{ syn},\sigma\rangle$. Specifically, the  initial state of the quantum walker is confined to $n_1=0$, and independent of synthetic momenta. Upon Fourier transformation to a full coordinate representation, $|n_1,k_\mathrm{syn},\sigma\rangle \to |n_1,n_\mathrm{syn},\sigma\rangle\equiv |\mathbf{n},\sigma\rangle$ 
this translates to localization at $|\bs 0,\sigma\rangle$ in both physical and synthetic space. 
The spreading of the quantum walker is thus given by  (see further details in Appendix~\ref{appen_observ}) 
\begin{align}
    \label{cf_2}
        \langle\Delta X^2 \rangle
         &\equiv 
    \sum_{\bs n}\sum_{\sigma',\sigma} n_1^2
   \overline{ |\langle \bs n,\sigma' | \mathcal{U}_F^t |\bs 0,\sigma\rangle|^2 }.
\end{align}
The correlation function 
Eq.~\eqref{cf_2}  describes the width in the physical $n_1$-direction of the wave packet initially prepared at $n_1=0$, and
its finite time scaling  encodes information on the  quantum walk dynamics.

\subsection{Topological invariants}

All our topological FMs to be discussed below are characterized by integer valued invariants. These numbers afford two different interpretations:

\textit{Topological invariants and FM classification:} The first relates to a classification of FM phases in terms of the periodic table of Hamiltonian insulators\cite{higashikawa2019floquet}.  
Its idea is to map the translation-invariant part of the  Floquet operator $U_\textbf{k}$ onto a block off-diagonal `auxiliary' Hamiltonian
\begin{equation}
    H_\textbf{k}=\left(\begin{array}{cc}
    &U_\textbf{k}\cr U_\textbf{k}^\dagger &
\end{array}\right). 
\end{equation}
This Hamiltonian inherits the symmetries of $U_\textbf{k}$, but in addition possesses a `chiral' symmetry due to its off-diagonality; it belongs to a  symmetry class different from the class of the Floquet theory. For example, if the latter is in class A (just unitary), $H$ will be in class AIII (chiral, no further symmetries.) 
Bott periodicity then implies that a class A Floquet theory realizes a FM state in odd effective dimension $D=d+d_{\rm syn}$ if the associated   $D$-dimensional Hamiltonian  in class AIII is also topologically non trivial.  Further, the presence of topologically non trivial phases of the Hamiltonian theory is signaled by invariants mathematically identical to those constructible for the Floquet theory. For example, in the above case, these invariants are `winding numbers' defined by a unitary map from odd-dimensional Brillouin zones $\textbf{k}\mapsto U_\textbf{k}$ into the unitary group. These winding numbers classify class AIII insulating phases in odd dimensions and class A Floquet metallic phases in even dimensions.

\textit{Topological invariants and localization theory:} To understand this statement in more concrete terms, we note that our Floquet theories are categorically disordered or chaotic. Their physical description requires real space methods, as defined by the nonlinear $\sigma$-models of disordered conductors. In these theories  protection against the effects of Anderson localization, i.e. topological metallicity, is introduced via topological terms (see Eq.\eqref{topological_action}  for a concrete example). These terms take physical effect provided their coupling constants are not vanishing. Below, we will demonstrate in two concrete case studies that the momentum space invariants responsible for the `abstract' classification of topological FMs indeed feature as coupling constants in the topological field theories. In this way, they serve a double function in the classification and the localization theory of FMs. In the latter context, they protect topological FMs from developing a `mobility gap' and force them to remain metallic, including FMs in low dimensions which would otherwise show strong localization.

\section{Three dimensional topological Floquet metal FM$_{1+2_{\rm syn}}$}
\label{subsection_fm1+2}

The concept of synthetic dimensions is 
general and can be realized for a wide class of 
driven or kicked Floquet systems~\cite{dahlhaus2011quantum,edge2012metallic,tian2016emergence,martin2017topological,petrides2018six}. In the following, we introduce a specific realization in $1+2_\mathrm{syn}$ dimensions, physically equivalent to the surface of a four dimensional topological insulator in symmetry class ${\rm A}$ (aka 'four-dimensional quantum Hall insulator').

\subsection{Model}

We consider a one dimensional quantum walker, whose forward and backward hopping amplitudes are time-dependent matrices coupling to the internal degrees of freedom  (see Fig.~\ref{fig:chiral_quantum_walk}). 
In the notation of the previous section, its  time evolution  from one discrete time step, $t$, to the next is defined through
\begin{align}
\label{U_step_chiral}
\hat U_{t} &= \frac{1}{2}(\sigma_0+\bold r_t \cdot \bs \sigma )\otimes \hat T_+ + \frac{1}{2} (\sigma_0-\bold r_t \cdot \bs \sigma)\otimes \hat T_-, 
\end{align}
where the specific choice
\begin{equation}
\label{eq:bold_r_t}
\bold r_t = (\cos \varphi_{2,t} \sin \varphi_{3,t}, 
\sin \varphi_{2,t} |\sin \varphi_{3,t}|,  \cos \varphi_{3,t} ),
\end{equation} 
will be motivated momentarily, and the time dependence of the phase arguments is defined below Eq.\eqref{f}. Turning to the gauge-equivalent representation in terms of a Floquet operator acting in a space with one physical and two synthetic dimensions, we describe the dynamics through the Floquet operator $\hat{\mathcal{U}}_F=\hat U_{\bold k}\hat W_{\bold{n}}$ with $U_\textbf{k}=U_{t=0}$ and $\hat T_\pm = e^{\pm ik_1}$. For later reference, we note that the unitary $U_\textbf{k}$ affords different representations, each useful in its own right. First, it is straightforward to verify that
 \begin{align}
  \label{d_vector_four_component}
  \hat U_{\bold k} 
  &= \exp \left[ik_1 ({\bold r} (\bold k) \cdot \bs \sigma) \right], \\
  {\bold r} (k) &= (\cos k_2 \sin k_3, \sin k_2 |\sin k_3|, \cos k_3 ). \nonumber
\end{align}
Alternatively, we may represent the spin matrices as rotations acting upon a translation operator with $z$-axis polarization: $\hat U_{\bold k} = R_3(-k'_2)R_2(-k_3)
\hat T_C R_2(k_3)R_3(k'_2)$ with $\hat T_C = e^{ik_1\sigma_3}$, the momentum $k'_2 = k_2 \,{\rm sgn}\, k_3$, and
 spin rotation operators $R_j(\varphi_l)=\exp \left( i \varphi_l \sigma_j /2 \right)$.

 A crucial feature of this realization is its non-analytic dependence on the momentum variables through $|\sin k_3|$. In the Fourier conjugate representation it translates to long ranged hopping $(\mathcal U_F)_{n_3n'_3} \sim |n_3 - n'_3|^{-2}$ \footnote{For $n\neq \pm 1$, $\int_{-\pi}^\pi e^{ik_2 n}|\sin k_2| dk_2 = \frac{2}{n^2-1}((-1)^{n+1} -1)$}. At this point, the synthetic dimensions begin to play an essential role: power law hopping in physical dimensions is difficult to engineer. More importantly, the $|\sin k_3|$ non-analyticity is the essential resource allowing us to sidestep the fermion doubling theorem and to realize a synthetic topological metal.

\begin{figure*}
\centering
\includegraphics[width=8.2cm]{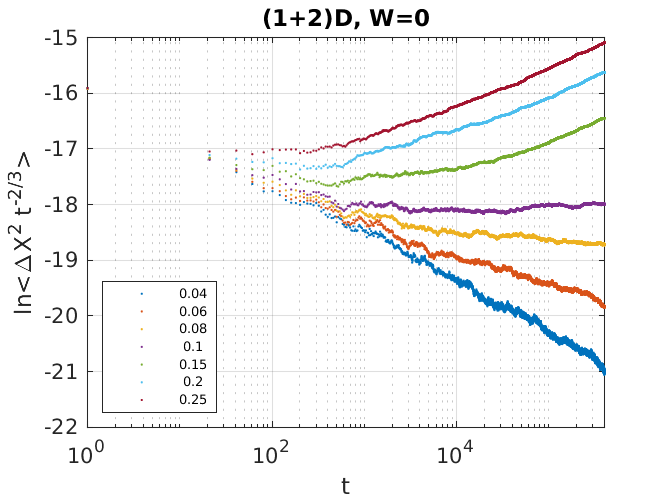}
\includegraphics[width=8.2cm]{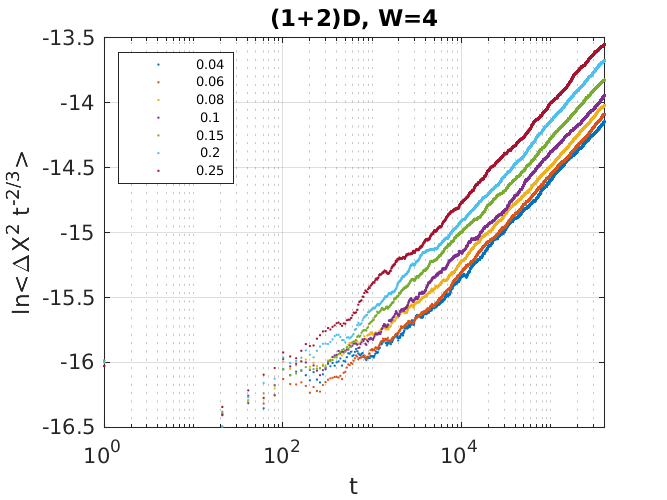}
\vspace{-0.5cm}
\caption{\label{fig:mit_3d} 
The spread of the quantum walker as a function of 
time steps for the 
two protocols and different values of bandwidth $w$ introduced in Eq.~\eqref{Uw}. 
To facilitate observation of the metal-to-insulator transition, 
the width of wave packet is rescaled by the critical scaling at the 
 transition, $\langle \Delta X^2 \rangle\sim t^{2/3}$.  
Left panel: The topologically trivial protocol, $W=0$,
 shows a metal-to-insulator transition for $w\simeq 0.1$. 
 Right panel: The topological non trivial variant, $W=4$,
shows no signatures of localization, 
i.e. $\langle \Delta X^2 \rangle\sim t$ for all values 
$w$ and up to $t\sim5\times 10^5$ time steps. 
}\end{figure*}

{\it Winding number:---}To elucidate this last point, we interpret Eq.~\eqref{d_vector_four_component} 
as a 
mapping from the $3d$ Brillouin zone torus to the two-dimensional 
special unitary group $\mathbb{T}^3\to {\rm SU}(2)$ 
(with unit determinant, $\det[U_{\bold k}]=1$),  
and assign the topological invariant 
\begin{align}
\label{eq:W_A_QKR}
W
&= 
\frac{1}{24\pi^2} \int d^3 \bold k\, \epsilon^{\mu\nu\rho}
\mathrm{tr}\left[
(U_{\bold k}^\dagger \partial_\mu U_{\bold k}) ( U_{\bold k}^\dagger \partial_\nu U_{\bold k} ) (U_{\bold k}^\dagger \partial_\rho U_{\bold k})  \right] 
\nonumber\\
&= 
\frac{1}{2\pi^2}\int d^3 \bold k \sin^2 k_1 |\sin k_3| 
= 4,  
\end{align}
where 
$\mu,\nu,\rho\in \{1,2,3\}$. 
As anticipated above, it is the non-analyticity 
$|\sin k_3|$  that 
leads to a non-vanishing winding number. 
Conversely, a model with analytic $k$-dependence would necessarily lead to a vanishing winding number, in accordance with the fermion doubling theorem.

 Driving protocols 
 with non-analytic functions, simulating power law hoppings in synthetic space,   
allow to 
construct Floquet operators with even winding numbers $W\in 2{\mathbb Z}$: The
starting point of our construction are realizations of unitary operators which display finite winding numbers over certain subsets of a
Brillouin zone, say $W = n$ in region ${\rm I}$ and 
$W = -n$ in region ${\rm II}$. (The numbers must add to zero by virtue of the
fermion doubling theorem.) The trick to generate a finite winding number then is to modify the momentum dependence in region ${\rm II}$ via a sign change in the momentum dependence which inverts the winding number to $W|{\rm II} = +n$ and
$W = 2n$ in total. 
We also constructed an alternative model with the minimal winding  possible in this scheme, $W=2$. Since it was rather involved  we opted, however, for the discussion of the simpler   $W=4$ variant, Eqs.~\eqref{U_step_chiral} and~\eqref{eq:bold_r_t}. We also emphasize that winding numbers for topological Floquet systems cannot be simply understood from their low energy effective Hamiltonians~\cite{Rudner:2013}.  $W=4$ of 
Eq.~\eqref{d_vector_four_component} is e.g. not related to four Weyl cones in a low energy description, but rather stores information on the entire Brillouin zone (see also discussion on dispersion of Eq.~\eqref{d_vector_four_component} in the next subsection).

The invariant Eq.\eqref{eq:W_A_QKR} indicates topological metallicity of our Floquet system. Within the  
alternative interpretation discussed above,  
the non-vanishing winding number, $W$, signals topological non-triviality of the `auxiliary' class ${\rm A III}$ Hamiltonian. 
Bott periodicity implies 
non trivial phases of $4d$ class ${\rm A}$ systems  (the four-dimensional quantum Hall effect), and the original Floquet system describes the physics of its three dimensional metallic surface state. 
In more concrete terms, we will see in section~\ref{subsection_low_energy_physics} that the winding number $W$ features as a building block in our construction of a gapless effective field theory equivalent to that of a three-dimensional topological metal.

\subsection{ Numerical simulations}
\label{subsection_numerical_similations}

To independently verify the topological metallic nature of 
the $3d$ dynamics simulated by the protocol Eq.~\eqref{U_step_chiral} 
 we have run numerical simulations. 
More specifically, we have simulated the quantum walk with trivial and non trivial winding numbers for varying effective disorder strengths by introducing additional bandwidths in the model, as we discuss next. 
This allows to test our main prediction, 
that is, the absence of Anderson localization for all disorder strength 
for finite windings $W$, contrasting Anderson localization at large disorder
for vanishing winding number $W=0$.

{\it Simulation details:---}We numerically study the time evolution 
of an initially localized wave packets, under influence of the $1d$ quantum 
walk operator Eq.~\eqref{U_step_chiral}.  
To allow for a comparison of topologically trivial and non trivial quantum walks 
with same energy-momentum dispersion relation of the clean system, 
we implement the walk with $|\sin \varphi_{3,t}|$, as indicated in Eq.~\eqref{d_vector_four_component},
and a second protocol with $|\sin \varphi_{3,t}|$ replaced by $\sin \varphi_{3,t}$. 
A metal-to-insulator transition with increasing disorder strength 
is expected for the second protocol. The static spatial disorder $\hat  U_{\rm dis}$ is implemented by
randomly drawing spin rotation matrices from the uniform Haar measure. 
 That is, the disorder strength is fixed
and we need to introduce some tunable  
 parameter allowing to drive a (possible) 
 metal-to-insulator transition.
We then notice that  Eq.~\eqref{d_vector_four_component} and  its topologically trivial cousin have no energy dispersion in $k_{2,3}$ direction, which 
 makes the latter always prone to localization. 
At the same time, we can perturb the original models to generate a dispersion with tunable bandwidth $w$ in $k_{2,3}$-direction. This then allows to study a delocalization transition as a function of  $w$. 
To realize  this idea, we multiply
the original single time-step evolution operators by the unitary operator
\begin{align}
\hat U_w 
&= 
\exp\left[ i w(\sin  k_{2,t} \sigma_1 + \sin k_{3,t} \sigma_2 ) \right],\label{Uw} 
\end{align}
which does the job.
We then study the time scaling of the average spread $\langle \Delta X^2 \rangle$
of the initial wave packet in the physical dimension.

{\it Results:---}Fig.~\ref{fig:mit_3d} shows the time evolution of 
$\langle \Delta X^2 \rangle$
for the two protocols and  
different values of the bandwidth $w$, as indicated in the legend.  
The vertical axis is rescaled by the 
time-dependence  $\langle \Delta X^2 \rangle \sim t^{2/3}$ expected at the metal-to-insulator transition~\cite{lemarie2009observation}. 
For the trivial protocol $W=0$ (left panel) one clearly sees the metallic ($w>0.1$) and insulating ($w<0.1$) regimes separated by 
critical scaling at $w\simeq0.1$. In contrast, the topological non trivial protocol with $W=4$ (right panel)
shows metallic behavior for all values of $w$, with $\langle \Delta X^2 \rangle \sim t$ 
up to the largest time steps $5\times 10^5$. 
The two incommensurate frequencies here were chosen as $\omega_2 = 2.4\sqrt 5$, 
$\omega_3 = 2.4\sqrt {15}$, and 
disorder averaging is over $~50$ realizations for each data point.  We also notice that the long-time numerical results are independent of the specific value for the incommensurate frequencies. 

\begin{figure*}
\centering
\includegraphics[width=18.2cm]{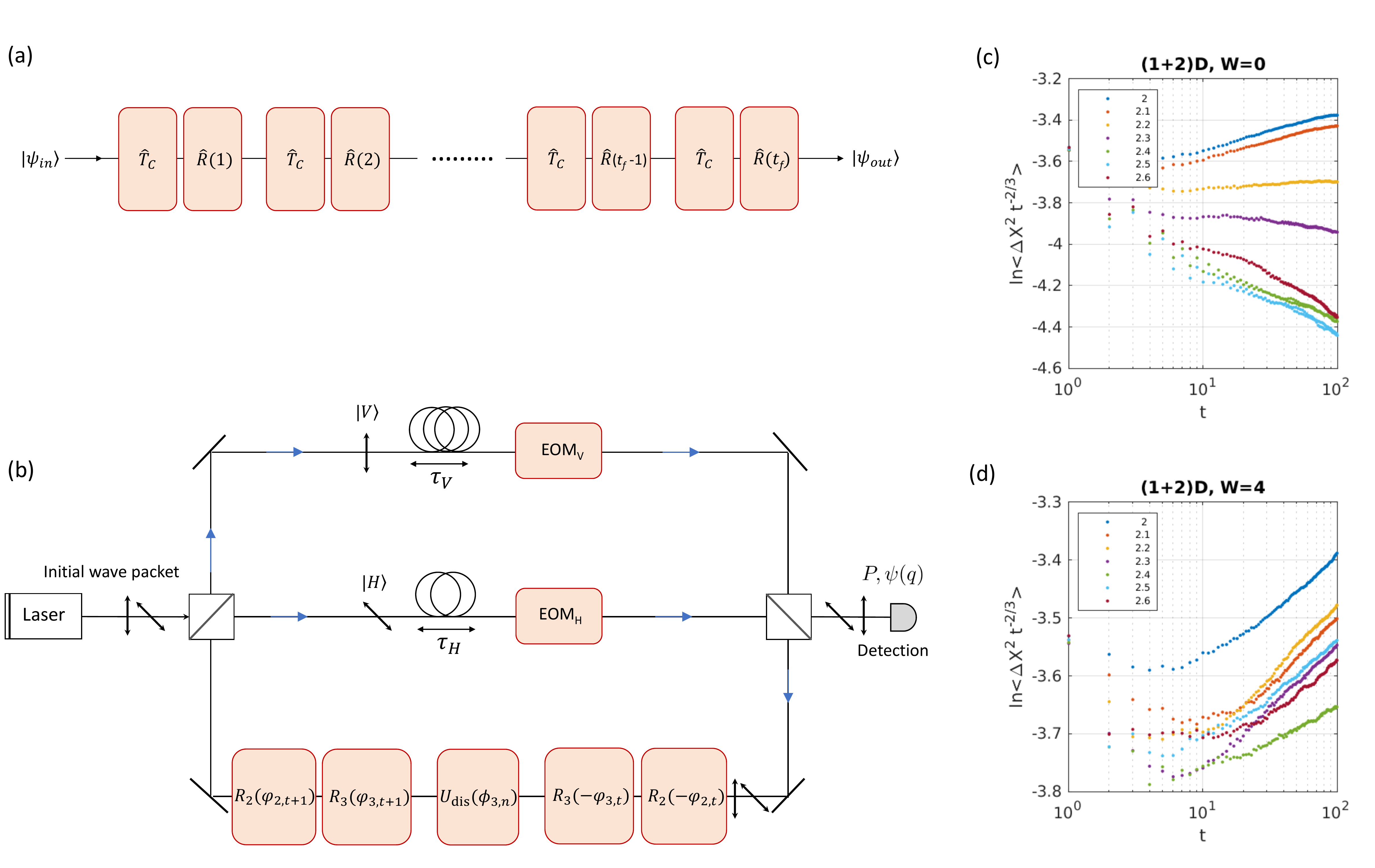}
\vspace{-0cm}
\caption{\label{fig:circuits} 
(a,b) Blueprint of an optical linear network simulating the topological 
surface states of a $4d$ quantum Hall insulator
in the quantum walk setting. 
The feedback loop is build of the step operator $\hat T_{\cal C}$ (upper and middle arms) 
 followed by the coin operator $\hat R(t)=R_{t+1} \hat U_{\text{dis}} R^\dagger_t$ 
(lower arm), for details see main text. 
To the left and right of the loop, source and detection units are connected.
The source consist of a laser  
and a polarizing beam splitter (PBS) that allows for the preparation of the initial state.
The passage to the detection unit can be activated by the dynamically tunable EOMs 
following the fiber lines of the step operation. In the detection unit 
 photons are registered by avalanche photo-diode (APD). 
(c,d) Numerical simulations of 
 wave packet spreading for trivial and topological quantum walks,
 here  for
 varying 
 incommensurate frequencies. 
The latter are chosen  as 
$\omega_2 = C \sqrt{5}$ and $\omega_3 = C \sqrt{15}$ for 	
 several values $2.0 \leq C \leq 2.6$ (see legends). 
Differences between the two systems become visible already after $t\gtrsim 10$ time steps:
 the  topological metal (d) shows robust diffusion 
 for all values of $C$, while 
 a crossover from diffusive to subdiffusive  dynamics is observed 
  as $C>2.4$ for the trivial metal (c). 
 \label{circuit_A}
}
\end{figure*}

In case of the trivial protocol $W=0$, deviations from classical scaling $\langle \Delta X^2 \rangle \sim t$   
signals Anderson localization, and eventually 
the dynamics will entirely freeze, $\langle \Delta X^2 \rangle \sim t^0$, 
on longer time and length scales. 
The important observation for us is that a clear distinction between quantum simulators
of trivial and topological metallic phases are noticeable  already for a small number of 
${\cal O}(20)$ time steps, see Fig.~\ref{fig:circuits}(c) and (d).  
This is crucial, once it comes to an experimental implementation of the quantum walks, as we discuss next.

\subsection{ Experimental realization: A FM$_{1+2_{\rm syn}}$ simulator}
\label{subsection_linear_optical_networks}

So far we have shown how the freedom of choosing operators 
of arbitrary $k$-dependence in the synthetic momentum space allows 
for the engineering of topological Floquet operators 
that cannot exist in autonomous lattice environments. 
Specifically, we have (i) proposed a concrete dynamical protocol based on a $1d$ quantum walk,
(ii) shown that this simulates a topological metal 
e.g. realized on the isolated surface of a $4d$ quantum Hall insulator, 
 and (iii) demonstrated that its most characteristic feature---absence of Anderson localization 
 at strong disorder---can be observed already after ${\cal O}(20)$ time steps. 
The final piece of our proposal is to indicate an experimental platform
that offers the required flexibility to implement dynamical protocols for spin-1/2 walkers.  
We here argue that linear optical networks are ideally suited to realize 
the proposed quantum simulators. 
After a brief review of their principal elements, we suggest 
a blueprint for the quantum Hall simulator.

{\it Principle elements:---}In the typical optical realization of a quantum walk, 
photons propagate through a network of linear elements, viz. 
beam splitters, phase shifters and polarization plates, realizing  
the ``step'' and ``coin'' operations. 
The step operation typically implements chiral hopping 
(here as a matrix in spin space) 
\begin{align}
\label{translation_chiral}
\hat T_{\cal C}
\equiv
\begin{pmatrix}
\hat T_+ &
\\
& \hat T_-
\end{pmatrix}, 
\end{align}
 translating the walker
to the right ($T_+$) or ($T_-$) left according to its spin 
being in the up (first component), respectively,
down state (second component).
The dynamical coin operations
  realize spin rotations, 
$\hat R(t)$. 
Using an Euler angle decomposition they can be 
generated from repeated application of 
 elementary rotations around any two of the three internal axes 
\begin{align}
\label{r_x}
\hat R_j(\varphi_{l,t}) 
&= 
\exp \left(i\varphi_{l,t}\sigma_j/2\right),
\quad
j=y,z, \quad l=2,3,
\end{align}
with Pauli matrices operating in spin space. 
The great flexibility offered by optical linear networks is that 
 Euler angles $\varphi_{l,t}=k_l+\omega_lt$ can be changed dynamically during  the realization of the quantum walk.

The successive application of 
chiral step and coin operations 
composing the quantum walk protocol 
is implemented in 
a ``feedback loop'', see Fig.~\ref{circuit_A}(b). 
Typically, a coherent laser pulse attenuated to an average single-photon per pulse 
injects photons into the linear network. 
Horizontal and vertical polarizations of the photon 
constitute 
the internal ``spin''-states. 
The step operation $\hat T_{\cal C}$ is realized in time, 
employing a polarizing beam splitter in combination with fiber delay lines.  
That is, horizontally and 
vertically polarized photons are separated by the beam splitter  
and send through fiber lines of different lengths. 
The length mismatch of the fibers introduces
 a well-defined delay between the two polarization components. When  
coherently recombined, the temporal separation of the two components is equivalent 
 to the spatial separation by two lattice sites  
 induced by the chiral translation to left and right neighbors of the $1d$ lattice. 
Dynamical coin operations $\hat R_j(\varphi_{l,t})$ 
are realized via tunable polarization rotations.
In practice, the dynamical control over only one rotation-axis, e.g. the $z$-axis is required, 
and rotations around remaining axes are realized 
by the combination 
with suitable polarization controllers, i.e. 
half- and quarter-wave plates~\footnote{\label{Basis_change}
For example, rotation around y-axis, $\hat R_2(\varphi) = e^{i\pi \sigma_1/4}\hat R_3(\varphi) e^{-i\pi \sigma_1/4}$, 
is implemented using the quarter-wave plates aligned at an angle $\pi/4$, 
$\hat C_{\text{QWP}}(\pi/4) = e^{-i\pi \sigma_1/4} $ \cite{geraldi2021transient}. 
While the rotation around x-axis, $\hat R_1(\varphi) = e^{-i\pi \sigma_2/4}\hat R_3(\varphi) 
e^{i\pi \sigma_2/4}$, is implemented by using the two half wave plates: 
$\hat C_{\text{HWP}}(\pi/8) \hat C_{\text{HWP}}(0) = e^{-i\pi \sigma_2/4}$~\cite{schreiber2010photons}.}.

The dynamical control is achieved via control voltages   
applied to fast-switching electro-optic modulators (EOMs) 
that change rotation-angles 
on time scales shorter than a step operation. 
 Recent progress 
allows to operate the latter without high additional losses 
and walks up to $t=30-40$ time-steps have been reported within this set up, 
see e.g. Refs.~\cite{schreiber2010photons,geraldi2021transient}. 
From the numerical simulations of the previous section,  
we expect this to be sufficient to distinguish the dynamics of 
a topological from a trivial Floquet metal. 
After the light pulses have been fed back into the 
loop of step and coin operations, realizing a single time step operation,  
for the desired number of time steps 
 they are released 
 to the detection unit, see  Fig.~\ref{circuit_A}(b). 
 Repeating the procedure for varying numbers of time-steps 
 and different realizations of 
 coin operations, one obtains  
  the walker's probability distribution which
  allows for the full characterization of its dynamics.  
 
{\it Blueprint for the FM$_{1+2_{\rm syn}}$:---}A detailed blueprint for an optical 
linear network simulating the topological 
FM$_{1+2_{\rm syn}}$ is shown in Fig.~\ref{fig:circuits}. A crucial observation here is that
the chiral translational $U_t$ (\ref{U_step_chiral})
with tunable spin-quantization axis $\bold r_t$ can be implemented via a
step operation $T_{\cal C}$ dressed by coin operations, $\hat U_t = \hat R^\dagger_t \hat T_{\cal C} \hat R_t$,
with the coin matrix $\hat R_t$ being a product of two elementary rotations, $\hat R_t
=
\hat R_y(\varphi_{3,t})
\hat R_z(\varphi_{2,t})$, 
where 
\begin{align}
\varphi_{3,t}
&=
k_3 +\omega_3 t,
\\
\varphi_{2,t}
&=
\begin{cases}
k_2+ \omega_2t,  
\quad \,
\sin(\varphi_{3,t})\geq0,
\\
-k_2 - \omega_2t,  
\quad \,
\sin( \varphi_{3,t})<0,
\end{cases}
\end{align}
(note the conditional value of $\varphi_{2,t}$ depending on the sign of $\sin(\varphi_{3,t})$, which follows from the definition~(\ref{eq:bold_r_t}) of the vector $\bold r_t$).
The role of $\hat R_t$ is to rotate $\bold z$-axis into the instantaneous spin quantization axis $\bold{r}_t$.

\begin{figure*}[t]
	\centering
	\includegraphics[width=14cm]{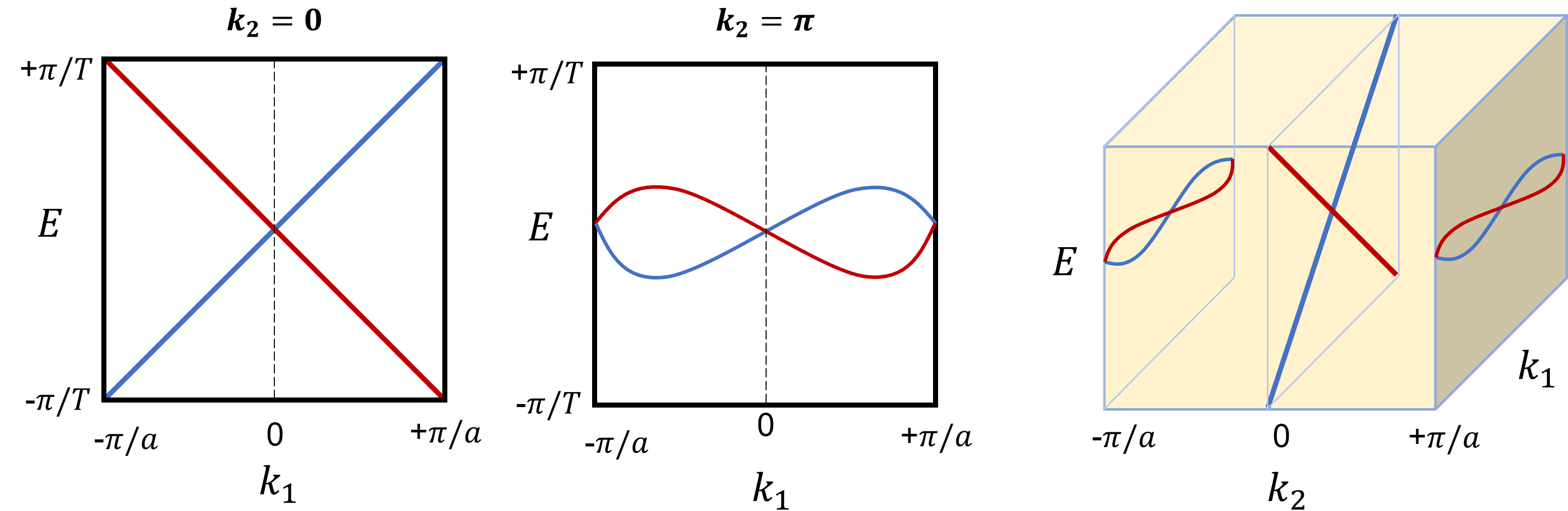}
	\vspace{-0cm}
	\caption{\label{fig:z2inv} 
	Dispersion relations  of the quasi-energy spectrum of the 
	two-band model along the high symmetry lines $k_2=0$ (left)
	and $k_2=\pi$ (middle), respectively, see also discussion in the main text.
	 The former contributes  to the topological invariant by the factor  
	  $\text{Pf}\left[ w_{\bold \Lambda_1} \right] \text{Pf}\left[ w_{\bold \Lambda_2} \right]=-1$,  while the latter gives the factor $\text{Pf}\left[ w_{\bold \Lambda_3} \right] \text{Pf}\left[ w_{\bold \Lambda_4} \right]=1$. Here 
	$w_\bold{k}$ is the antisymmetric sewing matrix. Right panel: Visualization of dispersions along the high-symmetry lines in the $2d$ Brillouin zone.}
\end{figure*}

In the linear network set-up it is convenient to start the feedback loop with a step operation.
Reorganizing thus spin rotations and  the local disorder potential in the originally defined one step evolution operator
$\mathcal U_{t,t-1}=\hat U _t \hat U_{\rm dis}$ (with $\hat U_{\text{dis}}$ specified below),
we construct the equivalent one as the following succession of step and coin operations, 
\begin{align}
\label{bp_a}
\mathcal U_{t+1,t}
&= 
\hat R(t)
\hat T_{\cal C},
\qquad
\hat R(t) 
=
\hat R_{t+1} \hat U_{\text{dis}} \hat R^\dagger_{t}.
\end{align}
This sequence is then iterated for the desired number of time steps. 

Fig.~\ref{fig:circuits}(a) schematically shows the elements of quantum walk operations to be applied to an initial localized wave packet before the detection after $t_f$ time steps. 
The actual implementation of the linear optical network can be prepared as in Fig.~\ref{fig:circuits}(b), realizing chiral quantum walk and coin operators.  The EOM$_{V,H}$  are equipped for the initiation and the readout of the quantum walk simulation.   
For the static disorder in real space, we suggest to follow the protocol used in the numerical simulations 
with fixed bandwidth. 
That is, choosing $\hat U_{\text{dis}}=R_z(\phi_{n_1})$ 
with static local angles $\phi_{n_1}$, randomly drawn from the unit circle $-\pi\leq \phi_{n_1} < \pi$, 
and frequencies $\omega_2$, $\omega_3$  
indicated in the previous section. 

Fig.~\ref{fig:circuits}(c) and (d) shows  numerical results for the quantum simulators 
of the
trivial ($W=0$) and topological ($W=4$) metal
 over a range of
experimentally accessible time steps.
Here
the bandwidth is set to $w=0$, and
incommensurate frequencies are varied as  
$\omega_2 = C\sqrt{5}, \omega_3 = C\sqrt{15}$ with values of $C$ indicated in the legend.
There are notable differences between the two systems already after $t\gtrsim 10$ time steps:
 Dynamics for the topological metal is diffusive for all values of $C$, while
the the trivial system 
shows a $C$-dependent behavior reminiscent of a metal-insulator transition as $C$ is increased.
Notice, however, that  the metallic behavior
for small values $C=2,2.1$ only holds for short time, and localization sets in at longer times
(i.e. for the trivial protocol there is no true metallic phase at $w=0$, as discussed above).

This finalizes our discussion of a quantum simulator for the surface states of a $4d$ quantum Hall 
insulator. 
We next discuss generalizations to other dimensions and symmetry classes.

\section{Two dimensional Floquet topological metal $FM_{1+1_{\rm syn}}$}
\label{subsection_simulator_quantum_spin_hall}

\begin{figure*}
\centering
\includegraphics[width=5.9cm]{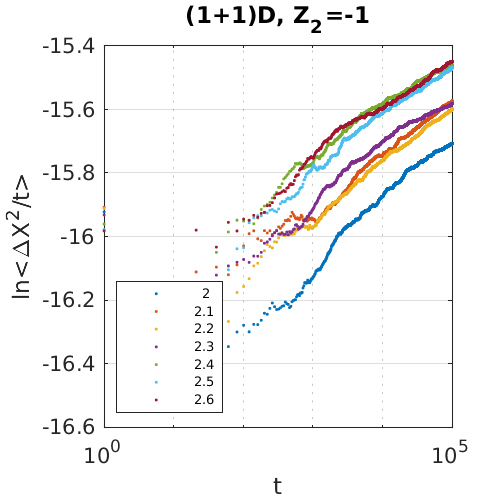}
\includegraphics[width=5.9cm]{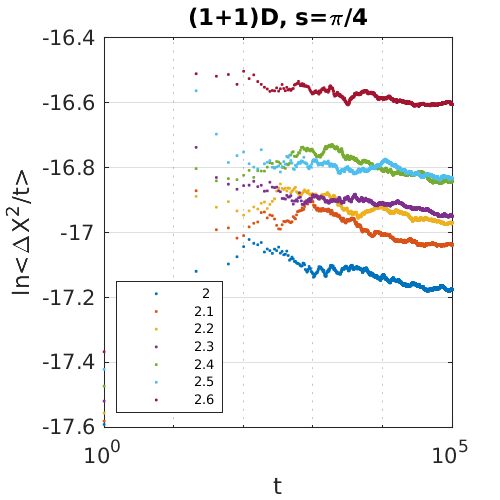}
\includegraphics[width=5.9cm]{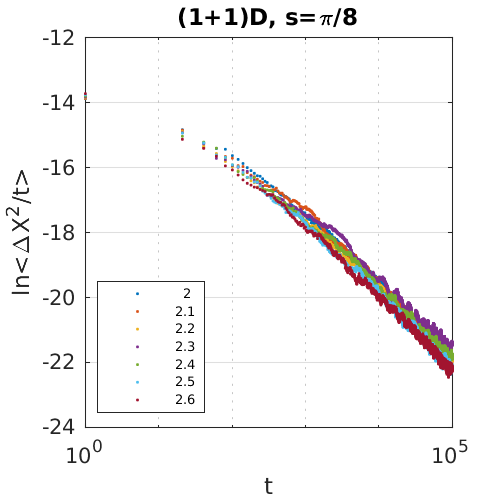}
\vspace{-0.5cm}
\caption{\label{fig:tfm_2d} 
The scaled width of a wave packet $\langle \Delta X^2 /t\rangle$ for different quantum walks in 
(1+1$_{\rm syn}$)D 
and monitored over $5 \times 10^5$ time steps. 
(Left) Topological quantum walk showing anomalous super diffusion, (middle) critical quantum walk of 
a class ${\rm A}$ model fine tuned to a topological phase transition, (right) topologically trivial quantum walk  subject to Anderson localization. 
 Colors correspond to incommensurate frequencies $\omega_2= \sqrt 5 C$ 
 with values $C$ indicated in the legend.
}
\end{figure*}

To illustrate the generality of our approach, we next discuss the example of a quantum simulator for surface states in a symmetry class different from the quantum Hall insulators. Specifically, we propose a simulator for the $2d$ surface states of a class ${\rm AII}$ quantum spin Hall insulator in $d=3$. 

To this end, we start out from a $1d$ quantum walk, Eq.~\eqref{U_step_reduced},  
with dynamical protocol,
\begin{align}
\label{fm1+1}
\vec r_0 
&= \frac{1}{2} (1-\cos\varphi_{2,t},0,0, \sin \varphi_{2,t}), 
\nonumber
\\
\vec r_r 
&= \frac{1}{4} (-1-\cos\varphi_{2,t},0,0, \sin \varphi_{2,t}),  \nonumber 
\\
\vec r_i 
&= \frac{1}{4} (0,|\sin \varphi_{2,t}|, -1-\cos\varphi_{2,t},0),  
\end{align}
where $\varphi_{2,t}=k_2+\omega_2 t$ and the frequency $\omega_2$ incommensurate to $2\pi$. The three vectors Eq.~\eqref{fm1+1} are orthogonal to each other, 
and unitarity of the single time-step evolution operator follows from $|\vec r_0|=|\sin \varphi_{2,t}/2|$ and $|\vec r_{r,i}|=\frac{1}{2}|\cos \varphi_{2,t}/2|$,
see also Appendix~\ref{quantum_walk_operator}. 
Upon Fourier transform in the physical coordinate and gauge transformation to eliminate time dependence of the driving protocol, we arrive at
the  Floquet operator
\begin{align}
\label{fm1+1k}
\hat U_{\bold k} =  (\hat r_0 \cdot \vec\sigma) + (\vec r_+ \cdot \vec\sigma) e^{ik_1}  + (\vec r_- \cdot \vec\sigma) e^{-ik_1},
\end{align}
where $k_2 = \varphi_{2,t=0}$ and $\vec r_\pm = (\vec r_r \pm i \vec r_i)$. 
We notice that the specific choice of  the driving protocol 
has lead to the non-analytical $n_2$-component 
$\propto|\sin k_2|$. It is again this unusual dependence, 
impossible to realize on a lattice with finite range hopping, 
which allows us to sidestep the fermion doubling theorem.
 It is readily  verified that Eq.~\eqref{fm1+1k}
satisfies the time reversal relation 
$\sigma_2 \hat U_{\bold k}^T\sigma_2 =\hat U_{-\bold k}$
of class ${\rm AII}$ systems. The latter 
host topological insulating $\mathbb{Z}_2$
phases in $3d$, and thus topological metallic Floquet phases in $2d$.

\subsection{Topological invariant}
To demonstrate the topological 
nature of the protocol Eq.~\eqref{fm1+1},
we focus on the translational invariant part  
$U_\bold{k}$, and 
consider the latter as a map from the $2d$ 
Brillouin zone torus to the 
special unitary group 
$\mathbb{T}^2\to {\rm SU}(2)$ ($\det[ \hat U_\bold{k}] =1$).
Time reversal symmetry 
imposes 
that $n(\bold{k})\parallel e_0$  
at the four time-reversal invariant momenta, 
 $\bold{\Lambda}_1 = (0,0)$, $\bold{\Lambda}_2 = (0,\pi)$, 
 $\bold{\Lambda}_3 = (\pi,0)$, and 
$\bold{\Lambda}_4 = (\pi,\pi)$. 
At these points the so-called sewing matrix $w_\bold{k} =  - i \sigma_2\hat U^T_\bold{k}$ is anti-symmetric and the map $\hat U_\bold{k}$ 
is thus characterized by the $\mathbb{Z}_2$ topological 
index,
\begin{align}
\label{eq:z2_QKR}
W_{\mathbb{Z}_2}
&= \prod_{i=1}^4 {\rm Pf}[-i\sigma_2 \hat U^T_{\bold{\Lambda}_i}]
=-1,
\end{align}
where in the last identity we used that for
 Eq.~\eqref{fm1+1k} $\hat U_{\bold{\Lambda}_1} = \sigma_0$, while $\hat U_{\bold{\Lambda}_i}
 = - \sigma_0$ for $i=2,3,4$. Notice that the non-triviality of the index follows from  
the specific choice of the  $n_2$-component. 

Building on the alternative interpretation of the topological invariant discussed earlier, $W_{\mathbb{Z}_2}$ signals topological non-triviality of its auxiliary class ${\rm DIII}$ Hamiltonian~\cite{higashikawa2019floquet}. Indeed,
time-reversal symmetry of the latter is inherited from the Floquet operator, while block off-diagonal structure induces the additional chiral  structure. In this interpretation Eq.~\eqref{eq:z2_QKR} then encodes topological properties of the class ${\rm DIII}$ system in $2d$~ 
\cite{ryu2010topological}.

For a more intuitive interpretation 
of Eq.~(\ref{eq:z2_QKR}), we prove in Appendix~\ref{section:Z2}  that the
 Pfaffians can be expressed as
${\rm Pf}(w_{\bold{\Lambda}_j})= - \exp{(i\epsilon_{\bold{\Lambda}_j}) }$, where $\epsilon_{\bold{\Lambda}_j}$ are the quasi-energies
 of $U_{\bf k}$ at the time-reversal invariant momenta $\bold{\Lambda}_j$.
The $\mathds{Z}_2$ topological invariant thus affords the alternative representation
\begin{equation}
\label{eq:Z2_phases}
    W_{\mathbb{Z}_2} = \exp\bigl( i \sum_{j=1}^4 \epsilon_{\Lambda_j} \bigr),
\end{equation}
which has a simple intuitive visualization. To this end, consider the 
$1d$ dispersion relations
$E^0_\pm(k_1) = \epsilon_\pm(k_1, k_2=0)$ and 
$E_\pm^\pi(k_1) = \epsilon_\pm(k_1, k_2=\pi)$ 
of the quasi-energy spectrum $\epsilon_\pm(\bold{k})$
of the two-band model 
along 
the two high symmetry lines $k_2=0,\pi$, respectively. 
 As shown in Fig.~\ref{fig:z2inv}, 
bands $E^0_\pm(k_1)$ touch at
 $\bold{\Lambda}_1$ and are split by energy $2\pi$ at  $\bold{\Lambda}_2$. 
Bands $E^\pi_\pm(k_1)$, on the other hand, touch in both momenta $\bold{\Lambda}_3$ and $\bold{\Lambda}_4$. This different pattern of the dispersion along the two high-symmetry lines results in the negative topological index $W_{\mathbb{Z}_2}=-1$,
as formalized by  Eq.~(\ref{eq:Z2_phases}).

\subsection{Numerical simulations}

We simulate the time evolution 
of an initially localized wave packet  in (1+1$_{syn}$) dimensions   
for three different Floquet operators  (all involving maximal disorder in the real coordinate, viz. random Haar unitaries):
The first simulates the topological Floquet metal, described in Eq.~\eqref{fm1+1}. Sharing the low energy physics of $2d$ class ${\rm AII}$ topological metallic surface states, we expect anomalous super-diffusion, $\langle \Delta X^2 \rangle \sim t \ln t$ \cite{tian2012anomalous}, 
which is confirmed in Fig.~\ref{fig:tfm_2d} left panel. 
Numerical calculations are performed for the incommensurate frequency $\omega_2=\sqrt 5 C$ (where the value $C$ is indicated in the legend), and each data point is obtained from averaging over 50 disorder realizations. 
The second Floquet operator simulates a critical state in class ${\rm A}$. That is, 
replacing the non analytic function in Eq.~\eqref{fm1+1} by an analytic function, $|\sin \varphi_{2,t}| \rightarrow \sin \varphi_{2,t}$, we obtain a  
(1+1$_{syn}$) dimensional class ${\rm A}$ model fine-tuned to  
a quantum critical point separating two topologically distinct  Anderson insulating phases~\cite{kim2020quantum}.  The presence of 
a  topological $\theta$-term fine-tuned to the angle $\theta=\pi$  in this case  
 protects against Anderson localization. Fig.~\ref{fig:tfm_2d} middle panel indeed indicates 
 sub-diffusion on all accessible time scales in our numerics. Notice that the energy dispersions for the first and  second Floquet operator are identical, and differences in the dynamics therefore root in the different topological terms. 
For the third Floquet operator, we tune the second Floquet operator away from the quantum critical point. 
The low energy physics in this case has a
$\theta$-term, however, with topological angle $\theta$  detuned from the critical value (see also next section). 
At long distances/times we then expect conventional  Anderson insulating behavior, which is confirmed
in the right panel of Fig.~\ref{fig:tfm_2d}.

\subsection{Incommensurability and synthetic dimensions}
\label{sec:Incommensurability}

So far we have discussed idealized quantum walk protocols
with irrational driving frequencies and frequency ratios. However, in view of our proposed  experimental implementations 
a comment on rational  approximations is due. A driving frequency $\omega = 2\pi \frac{p}{q}$ generates a synthetic dimension of finite extension $\sim q$. If the dynamics on length scales $\lesssim q$ is diffusive, a dimensional crossover takes place on time scales comparable to the diffusion time associated to distance scales $\sim q$. On larger scales, the system behaves as if it lived in one dimension lower.
The precision by which frequencies have to be chosen  thus 
depends on the experimentally probed time scales: 
the above  crossover should remain invisible in that it  occurs on scales larger than
the above crossover scales. (The precise value of these scales depends on system specific  parameters, notably the effective diffusion constant.) 

\begin{figure}
    \centering
        \includegraphics[width=4.2cm]{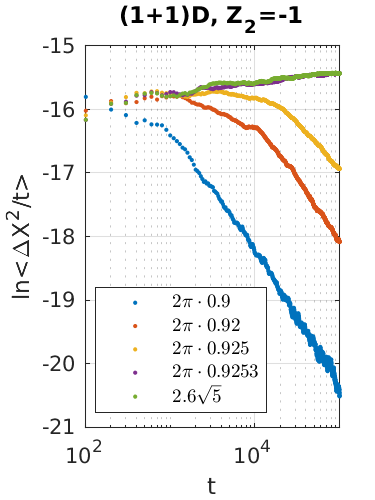}
        \includegraphics[width=4.2cm]{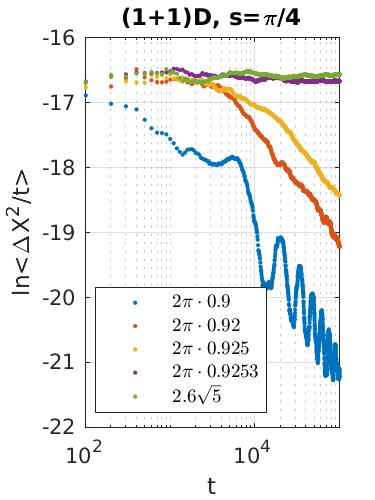}
    \vspace{-0cm}
    \caption{\label{fig:incommen} 
        Wave packet spread for quantum simulators of {\em finite size}
        class ${\rm AII}$ topological metals FM$_{1+1_{\rm syn}}$ (left), and $2d$ class ${\rm A}$ critical metals (right). The size of the compact dimension increases with the number of decimals kept in the rational approximation $\alpha$, and the dimensional crossover in the dynamics is observed at increasingly later times, see also discussion in main text.    
        }
\end{figure}

To make these general considerations more quantitative we 
numerically studied the protocol for the FM$_{1+1_{\rm syn}}$ 
of the previous section, substituting 
$\omega_2=2.6\sqrt{5}$ by 
$\omega_2 = 2\pi \alpha$  with rational approximations
of increasing periodicity
$\alpha=0.9,0.92, 0.925, 0.9253$. 
Left Fig.~\ref{fig:incommen} shows 
the width of a wave packet normalized by the 
width expected for diffusive dynamics,
$\Delta X^2/t$, as a function of $t$ on a $\log$-$\log$-scale. At the crossover scale to one-dimensional dynamics 
the (approximately) constant profile  for diffusive dynamics turns into a linear slope, characteristic
 for localized wave packets. As anticipated, the characteristic time scale increases with the periodicity of the rational approximation, i.e. the number of decimals kept in $\alpha$. 
Right Fig.~\ref{fig:incommen} shows 
the corresponding numerical results for the $2d$ critical class-A metal, also discussed in the previous section.

\section{Low energy physics}
\label{subsection_low_energy_physics}
\subsection{Class A}

To support our claim that the protocol Eq.~\eqref{U_step_chiral} 
simulates the isolated surface of a topological $4d$ quantum Hall insulator, 
we  next apply field theory methods for disordered systems.  
Our aim in this section is to show that the low energy physics of both systems is described 
by the same effective field theory.
Readers interested in more background material on field theories of disordered systems 
are invited to look into the supplemental material 
before or while plunging into this section.

To begin with, let us recall that the dispersion of the low energy excitations simulated by the 
clean contribution to the Floquet operator~\eqref{U_step_chiral}
is linear in $k_1$ and flat in all other directions. The emergence of Weyl 
fermions at low energies then results from the interplay of a non trivial topology 
and non-integrable chaotic fluctuations induced by the protocol. 
Indeed, we will see that the presence and stability of the Weyl fermions is topologically 
protected by the winding number, precisely in the same manner 
as the four Weyl cones of surface states in a $4d$ class ${\rm A}$ insulator at the $\nu=4$ integer quantum Hall plateau.

Applying methods for disordered systems~\cite{wegner1979,efetov1980zh, pruisken1982anderson, efetov1983kinetics, Pruisken:1984, Efetov-book},  
we can evaluate correlation functions 
for the dynamical quantum walk defined by \eqref{U_step_chiral} 
in the effective quantum field theory (QFT) framework. Its matrix degree of freedom $T$
acts in a replica space supplemented by additional causal (``retarded'' and ``advanced'') 
structure.
Within such QFT the physics at long time and length scales is described by an effective action
$S[T]=S_\sigma[T]+S_\mathrm{top}[T]$, consisting of two contributions 
\begin{align}
\label{diffusive_sigma_model_action}
S_\sigma 
&= \frac{1}{8}\sum_{i,j=1}^3
\sigma_{ij}^{(0)} 
\, 
{\rm Tr}\left(
\partial_i Q \partial_j Q 
\right)  ,
\\
\label{topological_action}
S_\mathrm{top}
&= 
iW \times  S_\mathrm{CS}[T].
\end{align}
Here the winding number $W$ is defined in  Eq.~(\ref{eq:W_A_QKR}), with $U$ the translational invariant part of the Floquet operator. 
The matrix field $Q=T Q_0  T^{-1}$, is expressed as rotations around
$Q_0\equiv \sigma_3\otimes \openone_R$ with Pauli matrix
$\sigma_3$ operating in the causal sector of 
the $2R$-dimensional vector space, 
and  `${\rm Tr}=\int d^3x\,{\rm tr}$' involves the trace 
over the latter and $3d$ space of physical and synthetical 
dimensions. 
Readers interested in further details are invited to look into 
the supplemental material 
where we explain the mathematical structures
and outline a derivation of the above action. We here restrict ourselves 
to a discussion of the physical implications 
of Eqs.~\eqref{diffusive_sigma_model_action},~\eqref{topological_action}.

The first observation is that $S_\sigma$ 
is the standard model for Anderson localization in disordered single particle systems, here in $3d$. 
For sufficiently strong disorder, viz. sufficiently small ``bare'' values $\sigma_{ij}^{(0)}$, the model   
flows to an Anderson insulating fixed point at long length scales
with vanishing coupling constant. 
We here consider Haar random disorder, for which 
\begin{equation}
\label{eq:sigma0_ij}	
\sigma_{ij}^{(0)}
=\frac{1}{2}
\int d^3\bold k\, \tr(\partial_{k_i} U_\bold{k}\partial_{k_j} U_\bold{k}^{-1})
\end{equation}
is purely determined by the translational invariant part of the Floquet operator.  
As we detailed above, we can then drive a metal-to-insulator transition by tuning 
the bandwidth of the system, see discussion around  Eq.~\eqref{Uw}.

A game changer to the Anderson insulating scenario at strong disorder 
is provided by the second, topological term with
Chern-Simons 
action $S_{\rm CS}$ 
\begin{align}
\label{chern_simons_action}
S_\mathrm{CS}
&=
\frac{1}{8\pi }\sum_{s=\pm }
s {\rm Tr}\left(A_s \wedge dA_s+\frac{2}{3}A_s \wedge A_s\wedge A_s\right).
\end{align} 
Here $A_s=T^{-1}d T P^s$,  where $P^s=\frac{1}{2}(1+s\sigma_3)\otimes \openone_R$  are projectors  
onto the retarded ($s=+$) and advanced  ($s=-$) sector of 
the $2R$-dimensional vector space.
The key observation then is that the 
combined action defined by Eqs.~\eqref{diffusive_sigma_model_action} and \eqref{topological_action} 
realizes a $3d$ topological metal, which unlike  systems with $W=0$ has a conductance growing in system size even at strong disorder. 
In the field theory language this means that the 
``bare'' coupling $\sigma_{ij}^{(0)}$ grows under renormalization and correlation functions show similar behavior. 
The same action, $S[T]=S_\sigma[T]+S_\mathrm{top}[T]$, 
has previously been
identified as describing 
(at length scales exceeding the mean free path)
$W$ disordered Weyl cones  
realized on the surface of a $4d$ class ${\rm A}$ quantum
Hall insulator surface~\cite{altland2016theory,zhao2015general}.  
We have thus established the equivalence between 
the protocol Eqs.~\eqref{U_step_chiral} for 
driven synthetic matter and quantum Hall insulator surface states,
demonstrating that both belong to the same universality class.

\subsection{class AII}

The non trivial $\mathbb{Z}_2$ 
index Eq.~\eqref{eq:z2_QKR} indicates that the 
low energy physics of the dynamical protocol Eq.~\eqref{fm1+1} is 
dominated by a single Weyl fermion,
similar to the isolated surface of a $3d$ quantum Spin Hall insulator. 
The single Weyl cone is not immediate 
from the low energy dispersion of the 
clean Floquet operator Eq.~\eqref{fm1+1}, but rather 
emerges as a consequence 
of the non trivial topology 
in combination with the 
chaotic fluctuations induced by the protocol. 

Applying field theory methods of disordered systems, we can again derive
a low energy effective theory $S[T]=S_{\rm \sigma}[T]+S_{\rm top}[T]$, 
that allows for the 
calculation of correlation functions 
at long time and length scales.
Here 
$S_{\rm \sigma}$ is the $\sigma$ model action, already introduced in Eq.~\eqref{diffusive_sigma_model_action}, 
now for the $2d$ system in the symplectic class ${\rm AII}$. The latter alone predicts 
a metal-to-insulator transition  
for strong disorder, a scenario changed by the second, topological contribution
\begin{align}
\label{top_a}
S_\mathrm{top}
&= 
i\frac{\theta}{\pi} \times  \left.\Gamma[g]\right|_{g(0,\bold x)=Q(\bold x)}.
\end{align}
Here $\theta=\pi W_{\mathbb{Z}_2}[U]$ is the topological angle from \eqref{eq:z2_QKR}, 
and
\begin{align}
\label{half_wess_zumino_witten}
\Gamma[Q]
&=
\frac{1}{24 \pi}  \int_{\cal M} \tr\left( \Phi_g \wedge \Phi_g\wedge \Phi_g \right)
\end{align} 
is ``half'' of a Wess-Zumino-Witten action, involving the usual deformation of the 
field degree of freedom $T$. Specifically,
$\Phi_g\equiv g^{-1}dg$ with $g(x_0=0,\bold x)=Q(\bold x)$, and integration is 
over half the $3$-torus 
${\cal M}= [0,1]\times [-1,1]^2 $.
This topological action was previously  identified~\cite{Ryu:2007, Koenig:2013} for the description of the $3d$ disordered quantum Hall insulator.  
 We refer the interested reader to 
the accompanying supplemental material for further explanations, 
and here only focus on a discussion of the physical implications.
These are similar to those of the Chern Simons action encountered 
in the unitary class. 
For $\theta = \pi$ 
the system flows to a conformally invariant 
quantum critical point, where the coupling constant of $S_\sigma$ 
assumes a disorder independent value~(\ref{eq:sigma0_ij}),   
implying the absence of Anderson localization also for strong disorder. 
The same principle of delocalization is at work on 
$2d$ surfaces of $3d$ 
topological spin quantum Hall insulators. The latter are 
indeed described by the same effective action~\cite{Ostrovsky:2007, Ryu:2007},
and we have thus shown that the protocol Eq.~\eqref{fm1+1} belongs to the same universality 
class as the isolated surface of a quantum spin Hall insulator.

\subsection{More on topological terms}
The possibility for disordered systems to escape the fate of Anderson localization
is signaled by topological terms in their low energy field theory 
description. 
Whether the latter are allowed depends on the dimension 
of the system and the target space of the field degree of freedom.
The effective action of the aforementioned $3d$ class ${\rm A}$  
system contains 
a Chern-Simons action, with coupling constant that is the winding number of the Floquet operator.
The winding number and Chern Simons action signal 
the presence of topologically inequivalent classes of mappings 
\begin{align}
\label{mapping_U_class_a}
U_{\bold k}: \,\,
&\mathbb{T}^3 
\mapsto 
{\rm SU}(2), 
\\
\label{mappting_T_class_a}
Q(x_0,\bold{x}): \, \,  
&\mathbb{T}^{(3+1)} 
\mapsto 
\mathrm{U}(2R)/[\mathrm{U}(R)\times \mathrm{U}(R)].
\end{align}  
Here  the boundary configuration $Q(x_0=0, \bold x)$ is 
parametrized by the 
$3d$ field 
$T(\bold x)$, used in the Chern-Simons action in Eq.\eqref{chern_simons_action}. 
Introducing the deformation parameter $0\leq x_0\leq 1$, continuously transforming the
boundary value $Q(x_0=0, \bold x)$ 
to the constant matrix $Q(x_0=1, \bold x )= \sigma_3\otimes\openone_R$, 
the Chern-Simons action can be 
expressed as a Wess-Zumino-Witten (WZW) term. The latter  
is precisely what is necessary for a $3d$ class ${\rm A}$ system 
to avoid Anderson localization (see 
supplemental material). 
The winding number of $U_{\bold k}$ given in \eqref{eq:W_A_QKR},
on the other hand, defines the coupling constant of the WZW term. 
That is, the combination of non-trivial homotopy groups
$\pi_3({\rm SU}(2))=\mathbb{Z}$ 
and
$\pi_4({\rm U}(2R)/[\mathrm{U}(R)\times \mathrm{U}(R)])=\mathbb Z$   
allows for non-vanishing coupling constants
weighting 
topologically non-trivial field configurations $T(\bold{x})$. 

For the $2d$ quantum spin Hall surface states 
it is the emergence of a WZW action, weighted by a $\mathbb{Z}_2$ topological angle
in the effective low energy description that allows for topological metallic phases. 
The involved maps in momentum- and real-space read 
\begin{align}
\label{mapping_U_class_aII} 
U_{\bold k}: \, \,
&\mathbb{T}^2 
\mapsto {\rm U}(2)/{\rm Sp}(2),
\\
\label{mapping_T_class_aII}
Q(\bold x):  \, \, 
&
\mathbb{T}^2 
\mapsto {\rm O}(4R)/[{\rm O}(2R)\times {\rm O}(2R)], 
\end{align}
with
non trivial homotopy groups, 
$\pi_2({\rm U}(2)/{\rm Sp}(2))=\mathbb{Z}_2$
and 
$\pi_2({\rm O}(4R)/[{\rm O}(2R) \times {\rm O}(2R)]) = \mathbb{Z}_2$, 
respectively. 
Both maps are thus characterized 
by non trivial $\mathbb{Z}_2$ indices, 
 and 
we already introduced a topological $\mathbb{Z}_2$ 
invariant for the Floquet operator Eq.~\eqref{mapping_U_class_aII}
in Eq.~\eqref{eq:z2_QKR}. 
Alternatively, one can express the $\mathbb{Z}_2$ index  
as ``half'' of a Wess-Zumino-Witten term, which
is readily extended to Eq.~\eqref{mapping_T_class_aII}, see also 
supplemental material for more details.

We conclude remarking that 
 the above discussion only relies on general structures, such as
the softmode manifold identified in the course of the construction of 
the effective field theory, 
and applies as a matter of principle. 
Whether there exist physical systems characterized by non-trivial couplings
 is an independent issue. The models we propose  in the earlier sections are one option how to realize 
 non trivial mappings utilizing the idea of engineered synthetic dimension. 
That is, in the present work we provide 
the field theories, physical models, and numerical confirmation of topological Floquet metals 
for both complex and real symmetry classes, which are
characterized by $\mathbb{Z}$ and $\mathbb{Z}_2$ topological indices, respectively.
The presented structures encompass topological Floquet metal in other dimension and symmetry classes.

\section{Discussion}
\label{section_discussion}

In this paper, we have introduced quantum simulators for 
topological surface states in isolation. 
Our proposal sidesteps the 
bulk boundary principle and overcomes the 
fermion doubling theorem, impeding the realization 
of isolated surface states in generic solid state (lattice) systems. 
The key element of our proposal is the 
dynamical generation of physical dimensions via external driving,  
using incommensurate frequencies. 
The  simulation of extra dimensions via driving physical
platforms has already been used in cold atom systems  
to measure the Anderson localization-delocalization
transition in three dimensions to a degree of resolution not reachable in solid state
materials. We here apply the idea to one-dimensional quantum walks 
of a spin-1/2 particle with time dependent spin rotation matrices, viz. ``coin operations'', 
following multi-frequency dynamical protocols.  
The latter provide 
a flexibility absent in lattice systems, which  allows for the simulation 
of (gauge) equivalent real-space dynamics involving long range hopping.  

We have illustrated the general idea on two specific examples, 
 the three-dimensional topological surface states of a four-dimensional 
quantum Hall insulator, and the two-dimensional surface states of a 
three-dimensional spin quantum Hall insulator. An inherent feature of 
both protocols is that the artificial generation of ``synthetic'' dimensions
induces diffusive dynamics in all (gauge) equivalent space directions 
after already a few iterations of the protocol, as verified 
in numerical simulations. 
Our approach, thus, simulates the surfaces of 
``disordered'' phases lacking translational invariance, 
which adds an element of realism.
 For both examples, we identified 
topological invariants 
showing the non trivial topological nature of the dynamical protocols. 
Comparing simulations of the 
latter to that of
topologically trivial parents with variable disorder strengths (respectively bandwidth)
clearly shows the impact of a non trivial topology. 
While  strong disorder turns the simulators of trivial metals into Anderson insulators, 
no signature of localization is found for the topological non trivial protocols for all disorder 
strengths, respectively, bandwidths. 
Importantly, the numerical simulations show
differences in the dynamics simulated by the different protocols 
already after an experimentally accessible number of $\sim{\cal O}(20)$ time steps. 
This also sets the precision to which frequencies have to be chosen in experiment. 
Approximating irrational numbers   
by rational  
generates {\it finite} 
rather than infinitely extended synthetic dimensions. 
As long as the corresponding diffusion time 
(i.e. the time required to explore the finite dimension)
exceeds the 
time scales probed in experiment, protocols with 
rational numbers can be used for all practical purposes.

Employing field theory methods, we have 
shown that the quantum simulators 
generate dynamics within the same universality class as the corresponding 
topological insulator surface states. 
Specifically, 
we demonstrated that 
the universal long-time dynamics of the 
 dynamical protocol is described by 
 precisely the same 
 topological field theory also proposed for 
 the simulated surface states. 
The field theory construction builds on the color-flavor transformation,  
and can be readily generalized to other symmetry classes and dimensions. 
Generalizing e.g. the simulator of topological quantum Hall surface states
to other (odd) dimensions, different from three, one can derive the  
  corresponding Chern Simons actions. 
  Similarly, a topological field theory with Chern Simons action 
can be derived for the simulator probing the surface states of 
a quantum spin Hall insulator 
  (`class ${\rm AII}$') in four dimensions, and different from the $\mathbb{Z}_2$ 
  field theory in three dimensions discussed here in detail. 
An exception is provided by class ${\rm AIII}$ systems.  
These cannot be simulated within the proposed scheme, since 
the gauge transformation, establishing 
the equivalence between the periodically 
driven and higher dimensional system, 
breaks chiral symmetry. 

Our proposal requires full dynamical control over a two-state internal degree of freedom (``spin''), 
which at the current state may be difficult to achieve in optical lattices. 
We, therefore, focused on the alternative platform of linear optical networks, 
similar to that used in Ref.~\cite{geraldi2021transient}. 
Specifically, time-multiplexing networks
with fast switching electro-optic modulators seem promising candidates for 
the implementation of the quantum simulators. 
We 
provided detailed blueprints 
for the experimental implementation of the two protocols
within existing set ups, 
realizing the quantum simulators of the surface states of 
a four dimensional quantum Hall insulator and  
a three dimensional quantum spin Hall insulator. 
We have shown in our numerical simulations 
that the experimental signature, 
viz. absence of Anderson localization, 
is observable within the experimentally 
realizable number of time steps. 
A tunable quantum simulator of topological surface states in isolation,
 would open fascinating experimental possibilities. 
 Specifically, it would 
 provide a new, direct window into the intriguing physics 
 resulting from the interplay of
 disorder and non trivial topology.

{\it Acknowledgments:---}T.~M.~acknowledges financial support by Brazilian agencies CNPq
and FAPERJ. K.W.K.~acknowledges financial support by Basic Science Research Program through the National Research Foundation of Korea (NRF) funded by the Ministry of Education (20211060) and Korea government(MSIT) (No.2020R1A5A1016518). A.A. and D.B. were funded by the Deutsche Forschungsgemeinschaft (DFG)
Projektnummer 277101999 TRR~183 (project A01/A03).

\begin{appendix}

\section{ Unitarity of quantum walk operator }
\label{quantum_walk_operator}

The general single time-step operator
Eq.~\eqref{U_step_generalization} simplifies to Eq.~\eqref{U_step_reduced} when focusing on quantum walks with short range hopping $m=\{-1,0,+1\}$. 
In momentum-representation, 
\beq
\hat U_{\bold k} &=& \sum_{m=0,\pm} (\vec r_m \cdot \vec \sigma) e^{im k_1}, \\
&=& \left[ \vec r_0 + (\vec r_++\vec r_-)\cos k_1 + i(\vec r_+-\vec r_-)\sin k_1 \right] \cdot \vec \sigma, \nonumber
\eeq
where $\vec r_m$ is a four-component vector and 
$\vec \sigma = (\sigma_0, i\bs \sigma)$. 
To satisfy unitarity, the vector multiplying $\vec\sigma$
 must be real valued, 
that is, $\vec r_+ = (\vec r_-)^*$. Expressing
$\vec r_+ =(\vec r_r + i \vec r_i)$ in terms of two real vectors $\vec r_r$, $\vec r_i$,
\beq
\!\!\!\hat U_{\bold k} = \left[ \vec r_0 +(\vec r_r + i \vec r_i)e^{ik_1}+ (\vec r_r - i \vec r_i)e^{-ik_1}\right] \cdot \vec \sigma, 
\eeq
and
requiring further that $\hat U_{\bold k} \hat U_{\bold k}^\dg = \mathds{1}$, 
the following relations can be verified 
\begin{align}
& |\vec r_{0}|^2 +|\vec r_{-}|^2 + |\vec r_{+}|^2 =1,\nonumber\\
&|\vec r_{0}|^2 +2|\vec r_{r}|^2 + 2 |\vec r_{i}|^2 =1,\nonumber\\
& \vec r_r \cdot \vec r_i =0, \vec r_{r}\cdot \vec r_0=0, \vec r_i \cdot \vec r_0 =0, \nonumber\\
\label{eq:Unitarity_cond}
&|\vec r_r| = |\vec r_i| = \frac{1}{2} \sqrt {1- |\vec r_0|^2}. 
\end{align}
These are stated below Eq.~(\ref{U_step_reduced}) in the main text.

\section{Spreading of a wave packet}
\label{appen_observ}
In this Appendix we demonstrate the equivalence of Eqs.~(\ref{cf} and (\ref{cf_2}). To this end we introduce the initial density matrix
\begin{equation}
\rho_0 = \frac{1}{N_{\rm syn}} \sum_{k_{\rm syn},\sigma}\ket{0,k_{\rm syn},\sigma} \bra{0,k_{\rm syn} , \sigma}, \quad \rho_0^2 = \rho_0,
\end{equation}
(here $n_1=0$ refers to the origin in the physical space and $N_{\rm syn} \gg 1$ is the number of initial phases) and note that Eqs.~(\ref{cf}) for $\Delta X^2$
can can be cast in the basis independent form
\begin{equation}
\langle \Delta X^2 \rangle={\rm tr} 
\overline{\left(\hat \rho_0  \mathcal{U}_{t,0}^\dagger \hat n_1^2 \mathcal{U}_{t,0}  \right)},
\end{equation}
where $\overline{(...)}$ refers to a disorder average.
The rationale behind this expression is the following. The average over initial phases (momenta $k_{\rm syn}$) implies the trace operation in the
extended Hilbert space, and we discretize the  corresponding momentum integral so that it becomes a sum over $N_{\rm syn}$ terms.  

Applying further the time-dependent gauge transformation introduced in section~\ref{subsection_synthetic_dimensions} one writes 
\begin{equation}
  \mathcal{U}_{t,0} = e^{ i t \sum_{j\geq 2 } \omega_j \hat n_j }\, \mathcal{U}_{F}^t,
\end{equation}
where the Floquet operator $\mathcal{U}_{F}$ was defined in Eq.~(\ref{eq:psi_t}). This ansatz gives us the equivalent expression for the width of a wave packet,
\begin{equation}
\langle \Delta X^2 \rangle = {\rm tr} \left(\hat \rho_0  (\mathcal{U}_{F}^\dagger)^t \, \hat n_1^2 \, \mathcal{U}_{F}^t  \right). 
\end{equation}
Lastly, to evaluate the trace above one can use a full coordinate representation, which gives us
\begin{align}
\langle \Delta X^2 \rangle & = \frac{1}{N_{\rm syn}}\sum_{\bold n\, \bold n',\sigma\sigma'} n_1^2 
\overline{|\bra{\bold n', \sigma'} \mathcal U_F^t \ket{\bold n, \sigma}|^2},
\end{align}
with $\ket{\bold n, \sigma} \equiv \ket{n_1 , n_{\rm syn}, \sigma}$ and $\ket{\bold n', \sigma} \equiv \ket{n_1 , n'_{\rm syn}, \sigma'}$.
We then notice that upon a disorder average the transition probability depends only on the difference in position, $\bold n'- \bold n$,
and thereby the expression~(\ref{cf_2}) in the main text is recovered.

\begin{figure*}
	\centering
	\includegraphics[width=18cm]{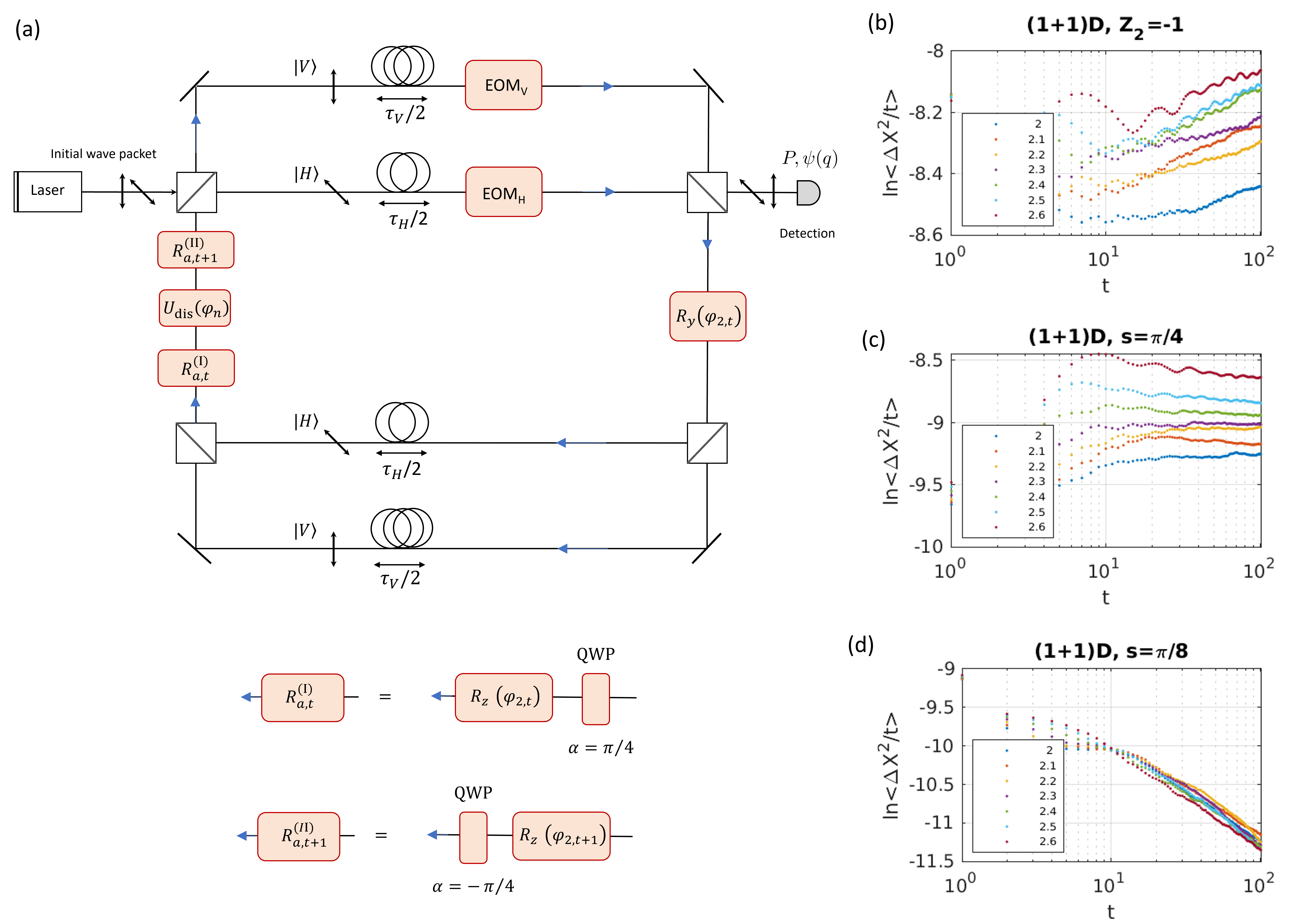}
	\vspace{-0cm}
	\caption{\label{fig:circuits_aii} 
		Blueprint of an optical linear network simulating the topological 
		surface states of a $3d$ quantum spin Hall insulator
		in the quantum walk setting. (a) 
		The feedback loop is build of the step operator $\hat T_{\cal C}^{\pm\frac{1}{2}}$ (upper and middle arms, respectively) 
		and the coin operators $\hat R_a$ and $\hat R_b$ 
		(right and left arm, respectively), for details see main text. 
		To the left and right of the loop, source and detection units are connected.
		(b) Numerical simulation of a topological metal ($\mathbb{Z}_2=-1$) in a $(1+1_{\rm syn})D$ quantum walk for short time steps accessible by experiments. The width of the wave packet scaled by time is plotted on a $\log$-scale, showing that diffusion is anomalously fast. (c) Critical quantum walk in $(1+1_{\rm syn})D$ class ${\rm A}$ at a topological quantum phase transition, showing the same scaling with the classical diffusion $\langle \Delta X^2 \rangle \sim t$. (d)  Quantum walk in $(1+1_{\rm syn})D$ class ${\rm A}$ without topological term, showing Anderson localization. Colors correspond to incommensurate frequencies $\omega_2= \sqrt 5 C$ 
		with values for $C$ as indicated in the legends.
	}
\end{figure*}

\section{ Blueprint for the FM$_{1+1_{\rm syn}}$ simulator }
\label{appendix_spin_qunatum_hall_insulator}

A detailed blueprint for the optical linear network 
 simulating the topological FM$_{1+1_{\rm syn}}$ 
 is shown in Fig.~\ref{fig:circuits_aii}. 
The dynamical protocol, Eq.~\eqref{fm1+1},
involving all three components $\hat R_{\pm,0}$
requires a more complex set-up in comparison to FM$_{1+2_{\rm syn}}$ in class A, which now has to be build 
from two chiral half-step and two coin operations. 
Therefore we start by summarizing the optical scheme in Fig.~\ref{fig:circuits_aii} and
then provide its justification.
To this end we decompose the translational invariant part 
of the single time step evolution 
operator, Eq.~\eqref{fm1+1}, into the product 
\begin{align}
\mathcal U_{t+1,t}
&=\hat U_{\rm dis}\hat R_a(t)\hat T_{\cal C}^{\frac{1}{2}} R_b(t) \hat T_{\cal C}^{\frac{1}{2}},  \label{eq:U_AII}
\end{align}
of a chiral half-step, $\hat T^{\frac{1}{2}}_{\cal C}$, 
and coin operators, $\hat R_{a/b}$. 
(The former
are positioned in the four horizontal 
arms, notice the half fiber lengths $\tau_H/2$ and $\tau_V/2$,
and the latter are placed in the vertical arms.) 
Dynamical EOMs after fiber lines of the first step operation 
 allow to terminate the walk by sending the photons to the detection unit. 
The coin operators are chosen as
\begin{equation} 
\hat  R_b(t) = Y R_z(-\varphi_{2,t}) Y^\dagger \equiv R_y(\varphi_{2,t}),
\end{equation}
where $Y= e^{i\pi \sigma_1/4}$
is the matrix of $y$-basis change~\footnotemark[\value{footnote}], and
$ 
\hat R_a(t) = \hat R_{a,t+1}^{\text{(II)}} \hat R_{a,t}^{\text{(I)}}$,
with
\begin{align}
\hat R_{a,t}^{\text{(I)}} 
&=
\begin{cases}
R_z(\varphi_{2,t})\hat Y^\dagger,
\quad \sin \varphi_{2,t} \geq 0, 
\\
\hat Y^\dagger, 
\quad \sin \varphi_{2,t} <0, 
\end{cases}\label{Ra1}
\end{align}
and 
\begin{align}
\hat R_{a,t+1}^{\text{(II)}} 
&=
\begin{cases}
\hat Y, 
\quad \sin \varphi_{2,t+1} \geq 0,
\\ 
\hat Y R_z(\varphi_{2,t+1}),
\quad \sin \varphi_{2,t+1} < 0, 
\end{cases}\label{Ra2}
\end{align}
with
$ \varphi_{2,t}=k_2 + \omega_2 t$. 
Finally,
disorder is introduced 
by placing in between
 $R_{a,t}^{\text{(I)}}$ and $R_{a,t+1}^{\text{(II)}}$
the local, time reversal invariant random potential  
$U(n_1)=e^{i\phi_{n_1} \sigma_0}$, with
 position dependent angles $\phi_{n_1}$, randomly 
drawn from the unit circle $-\pi\leq \phi_{n_1} < \pi$.

Coming back to the justification of~(\ref{eq:U_AII}) we note that expressing the time evolution operator in Eq.~\eqref{fm1+1} as the product 
of elementary chiral translations and coin operators is not immediately straightforward. To start with, we first notice that upon replacing $|\sin k_2| \rightarrow \sin k_2$ in Eq.\eqref{fm1+1k}, the model is reduced to the familiar $2d$ class ${\rm A}$ Floquet insulator~\cite{rudner2013anomalous, kim2020quantum}, which is built by the multiplication of simpler operators. 
For an implementation of the protocol with absolute value
$|\sin k_2|$,
we separate cases $\sin k_2 \leq 0$ and $\sin k_2 >0$. 
Specifically, a class ${\rm A}$ $2d$ Floquet model with homotopy parameter 
$s$ is written as~\cite{kim2020quantum} the product of four unitary operators $\hat U_{\text A} = \hat U_4\hat U_3\hat U_2\hat U_1$, with 
\begin{align*}
\hat U_i(k) &= \cos(s) + i\sin(s) \begin{pmatrix}
 & e^{-i\bold k \cdot \bold v_i} \\ e^{i\bold k\cdot \bold v_i} &
\end{pmatrix}.
\end{align*}
Here $(\bold v_1,\bold v_2,\bold v_3,\bold v_4) = (0,-\bold e_1,-\bold e_1+\bold e_2 , \bold e_2)$, 
with $\bold e_{1,2}$ lattice unit vectors in the horizontal/vertical direction. 
At $s=\pi/4$, the product of unitaries can be expanded in the quantum walk form
\begin{align*}
\hat U_{\text A} = (\hat r_0 \cdot \vec \sigma) + (\vec r_+ \cdot \vec\sigma)e^{ik_1}+(\vec r_- \cdot \vec\sigma)e^{-ik_1},
\end{align*}
where 
\begin{align*}
\vec r_0 &= \frac{1}{2} (1-\cos k_2, 0, 0, \sin k_2), \\
\vec r_r &= \frac{1}{4} (-1-\cos k_2,0,0, \sin k_2), \\
\vec r_i &= \frac{1}{4} (0, \sin k_2, -1-\cos k_2, 0). 
\end{align*}
Following the general recipe outlined in Sec.~\ref{subsection_synthetic_dimensions}, 
we simulate the $2d$ dynamics as a $1d$ quantum walk with time dependent protocol, 
replacing momentum $k_2 \rightarrow \varphi_{2,t}$ by a time dependent angle. 
We then notice that  for $\sin \varphi_{2,t}>0$  vectors $\vec r_{0,r,i}$ of $\hat U_{\text A}$ are identical 
to that of $\hat U_{\text {AII}}$ in Eq.\eqref{fm1+1}. Hence, when $\sin \varphi_{2,t}>0$, the quantum walk operator in Eq.\eqref{fm1+1} can be written as the product of the four unitary operators, $\hat U_{\text {AII}} (\sin\varphi_{2,t}>0) = \hat U_{\text A} (k_1,\varphi_{2,t})$. 
When $\sin \varphi_{2,t}<0$, 
on the other hand, one can verify that 
$\hat U_{\text{AII},\sin \varphi_{2,t}<0}(k_1,\varphi_{2,t}) = \hat U^T_{\text A} (-k_1,\varphi_{2,t},s=\frac{\pi}{4})$.

Next, we express $\hat U_{j=1,2,3,4}$ as a combination of shift and rotation operators,
\begin{align*}
\hat U_1 &= e^{is \sigma_1}, \\
\hat U_2 &= \hat T_C^{-\frac{1}{2}}e^{is \sigma_1}\hat T_C^{\frac{1}{2}},\\
\hat U_3 &= e^{i\varphi_{2,t}\sigma_3/2} \hat T_C^{-\frac{1}{2}} e^{is \sigma_1}\hat T_C^{\frac{1}{2}}e^{-i\varphi_{2,t}\sigma_3/2},\\
\hat U_4 &= e^{i\varphi_{2,t}\sigma_3/2}e^{is \sigma_1}e^{-i\varphi_{2,t}\sigma_3/2},
\end{align*}
with `half' translation operator $\hat T_C^{\frac{1}{2}} = e^{ik_1\sigma_3/2}$.
We stress that the full Floquet operator $\hat U_{\text A}$ is $2\pi$ periodic in $k_1$
and the appearance of a 
`half' translation operator
 does not imply a doubling of the unit cell. 
The same is true for the topological Floquet metal model. 

The $2d$ class ${\rm AII}$ model is expressed as
\begin{align}
\label{eq:U_AII_1}
&\hat U_{\text{AII},\sin \varphi_{2,t}\geq 0}(k_1,\varphi_{2,t})\\ 
&= \hat U_{\text A}(k_1,\varphi_{2,t},s=\frac{\pi}{4}) \nonumber\\
&=e^{i\varphi_{2,t}\sigma_3/2} Y\hat T_C^{-\frac{1}{2}} Y e^{-i\varphi_{2,t}\sigma_3/2}Y \hat T_C^{\frac{1}{2}}Y, \nonumber\\
&=-e^{i\varphi_{2,t}\sigma_3/2} Y^\dagger \hat T_C^{\frac{1}{2}} Y^\dagger e^{-i\varphi_{2,t}\sigma_3/2}Y \hat T_C^{\frac{1}{2}}Y, 
\nonumber
\end{align}
and to arrive at this result we used 
that $Y \hat T_C^{-\frac{1}{2}}Y = - Y^\dagger \hat T_C^{\frac{1}{2}} Y^\dagger$
and commutativity of operators $\hat T_C^{\frac{1}{2}}$ and $e^{-i\varphi_{2,t}\sigma_3/2}$.
On the other hand, 
\begin{align}
\label{eq:U_AII_2}
&\hat U_{\text{AII},\sin \varphi_{2,t}<0}(k_1,\varphi_{2,t})\\ 
&= \hat U^T_{\text A} (-k_1,\varphi_{2,t},s=\frac{\pi}{4}) \nonumber\\
&= Y \hat T_C^{-\frac{1}{2}} Y e^{-i\varphi_{2,t}\sigma_3/2} Y \hat T_C^{\frac{1}{2}}Y e^{i\varphi_{2,t}\sigma_3/2},\nonumber\\ 
&= -Y^\dagger \hat T_C^{\frac{1}{2}} Y^\dagger e^{-i\varphi_{2,t}\sigma_3/2} Y \hat T_C^{\frac{1}{2}}Y e^{i\varphi_{2,t}\sigma_3/2}, \nonumber
\end{align}
and from above relations~(\ref{eq:U_AII_1}) and (\ref{eq:U_AII_2}) we notice that 
$\hat U_{\rm AII} = R_a^{\rm (I)} \hat T_C^{\frac{1}{2}} R_b \hat T_C^{\frac{1}{2}} R_a^{\rm  (II)}$. 
Up to a cyclic permutation of the operator $R_a^{\rm (II)}$ this is equivalent to the clean part of
${\cal U}_{t+1,t}$, see Eq.~(\ref{eq:U_AII}).
Notice that in both cases 
$\hat U_{\text{AII}}$ involves the same unitary sandwiched between the two $\hat T_C^{\frac{1}{2}}$, which 
simplifies the implementation of $\hat R_b(t)=Y^\dagger e^{-i\varphi_{2,t}\sigma_3/2} Y$. 
$\hat R_a(t)$, on the other hand, 
depends on the sign of $\sin \varphi_{2,t}$.

\section{$\mathbb{Z}_2$ topological invariant}
\label{section:Z2}
In this Appendix we further discuss the $\mathbb{Z}_2$ topological invariant for the 2-band model and prove the relation~(\ref{eq:Z2_phases}).
The unitary operator $\hat U_{\bold k}$ 
gives rise to the auxiliary Hamiltonian \cite{roy2017periodic, higashikawa2019floquet}, 
\begin{align} \label{eq:Hu}
    \tilde H_U (\bold k) = \begin{pmatrix}
    0 & U_{\bold k} \\ U_{\bold k}^\dg & 0 
    \end{pmatrix},  
\end{align}
which shares time-reversal and particle-hole symmetries, 
$\hat \Theta_1  \tilde H_U (\bold k) \hat \Theta_1^{-1} =  \tilde H_U (-\bold k) $, 
and 
$\hat \Theta_2  \tilde H_U (\bold k) \hat \Theta_2^{-1} = - \tilde H_U (-\bold k) $,  
respectively,  
with $\hat \Theta_1 = \tau_1 \otimes i\sigma_2 \mathcal K$, and 
$\hat \Theta_2 = i\tau_2 \otimes i\sigma_2 \mathcal K $. 
That is, $H_U$ belongs to class ${\rm DIII}$. Notice that in Eq.\eqref{eq:Hu} off diagonal elements are the Floquet unitary operator without band flattening. This allows us to make a connection between the $\mathbb{Z}_2$ invariant and the eigen-energies at  time-reversal invariant momenta. 

A way to compute the $\mathbb{Z}_2$ topological invariant then is as follows~\cite{ryu2010topological}:  
The Hamiltonian has two valence bands at energy $E=-1$, and their eigenvectors are
\begin{align}
    u^-_{1}(\bold k) = \frac{1}{\sqrt{2}} \begin{pmatrix}
    -1 \\ 0 \\ U^*_{\bold k,11} \\ U^*_{\bold k,12} 
    \end{pmatrix}, \,\,\,\,\,
    u^-_{2}(\bold k) = \frac{1}{\sqrt{2}} \begin{pmatrix}
    0 \\-1 \\  U^*_{\bold k,21} \\ U^*_{\bold k,22} 
    \end{pmatrix}, 
\end{align}
where $U^*_{\bold k,ij} =(U^\dg_{\bold k})_{ji}$. The sewing matrix, needed to compute the topological invariant~\cite{ryu2010topological}, can be obtained from these two vectors as  
$(w_{\bold k})_{ab} = \dirac{u^-_{a}(-\bold k)}{\hat \Theta_1 u^-_{b}(\bold k) } $. That is,
\begin{align}
    w_{\bold k} &=\frac{1}{2} \begin{pmatrix}
    -U_{\bold k,12} + U_{-\bold k,12} & -U_{\bold k,22} - U_{-\bold k,11} \\
    U_{\bold k,11} + U_{-\bold k,22} & U_{\bold k,21} - U_{-\bold k,21}
    \end{pmatrix}, \\
    &=\begin{pmatrix}
    -U_{\bold k,12} & -U_{\bold k,22}\\
    U_{\bold k,11}  & U_{\bold k,21} 
    \end{pmatrix}, \\
    &= -i\sigma_2 U^T_{\bold k} , 
\end{align}
where in the second line time-reversal symmetry of the unitary operator was used, i.e. $U_{\bold k,11}=U_{-\bold k,22}$,  $U_{\bold k,12}=-U_{-\bold k,12}$ and $U_{\bold k,21}=-U_{-\bold k,21}$ 
(as follows from $\sigma_2 U_{\bold k} \sigma_2 = U_{-\bold k}^T$).
One can then readily verify that the sewing matrix is anti-symmetric at time-reversal invariant momenta $\bold \Lambda_1 = (0,0)$,  $\bold \Lambda_2 = (\pi,0)$,  $\bold \Lambda_3 = (0,\pi)$,  and $\bold \Lambda_4 = (\pi,\pi)$, i.e.  
\begin{align}
w_{\bold \Lambda_j} =\frac{1}{2} \begin{pmatrix}
    0 & -U_{\bold \Lambda_j,11} - U_{\bold \Lambda_j,22} \\
    U_{\bold \Lambda_j,11} + U_{\bold \Lambda_j,22} & 0
    \end{pmatrix}, 
\end{align}
where $U_{\bold \Lambda_j,11}=U_{\bold \Lambda_j,22}$, and the Pfaffian is 
$\text{Pf}\left[w_{\bold \Lambda_j}\right]=- \frac{1}{2} (U_{\bold \Lambda_j,11} + U_{\bold \Lambda_j,22}) = -\frac 1 2 \tr [U_{\Lambda_j}] = -\exp\left( i\epsilon_{\bold \Lambda_j}\right)$. 
Finally, the $\mathds{Z}_2$ topological invariant becomes 
\begin{align}
    W_{\mathbb{Z}_2} &= \prod_{j=1,2,3,4} \text{Pf}\left[w_{\bold \Lambda_j}\right]
     = \exp \left( i \sum_{j=1,2,3,4} \epsilon_{\Lambda_j} \right), 
\end{align}
which implies that the condition for a non trivial Floquet topological metal, $W_{\mathbb{Z}_2}=-1$,  translates  into  $\sum_{j=1,2,3,4} \epsilon_{\Lambda_j} = \pi$ (mod $2\pi$).

\end{appendix}

\bibliography{bibliography}

\pagebreak
\begin{center}
\textbf{\large Supplemental Materials}
\end{center}

\setcounter{section}{0}
\setcounter{equation}{0}
\setcounter{figure}{0}
\setcounter{table}{0}
\setcounter{page}{1}
\makeatletter
\renewcommand{\thesection}{S\arabic{section}}
\renewcommand{\theequation}{S\arabic{equation}}
\renewcommand{\thefigure}{S\arabic{figure}}

In this supplemental material we provide details on the effective field theory description of the quantum simulator protocols. 
We start out with a brief review of the standard effective field theory for  
disordered single particle systems, then outline how to map 
the protocols discussed in the main text to the latter, and finally deepen our discussion of topological terms.

\section{Field theory review }
\label{appendix_field_theory_details}

\subsection{Diffusive non-linear $\sigma$ model}
The quantum dynamics of disordered single 
particle systems at long length- and time-scales 
is described by a diffusive non-linear $\sigma$ model. 
The latter bears similarities to Ginzburg-Landau 
theories and encodes the physics  
of symmetry breaking and (critical) soft-mode fluctuations
related to Anderson localization. 
Different from the former, 
the field degree of freedom of the $\sigma$ model however 
does not afford the interpretation of 
an order parameter.
More specifically, the model is formulated in terms of a matrix degree of freedom 
$Q$ which satisfies the nonlinear constraint $Q^2=\openone$. 
In its
simplest replica variant the matrix is operating in a 
$2R$-dimensional vector space formed by  
$R$ replicas 
which exist in two ``causal'' variants, a ``retarded'' and an ``advanced''. The 
replica structure helps to overcome the notorious problem of disorder averaging the 
logarithm of the partition function,
which serves as a  ``generating function'' for observables. 
The causal structure is introduced 
to  generate the typical observables of interest (see Section~II B in the main text), 
viz. disordered averaged probabilities
(the product of retarded and advanced propagators), 
from a single generating function. 

Anderson localization can be viewed as the restoration of 
rotational symmetry in causal space. 
Indeed, the derivation of the field theory 
builds around a saddle point 
that describes the
disorder induced level broadening. 
This is isotropic in replica space, while
causality breaks rotational symmetry in retarded and advanced 
components. 
Formally, 
$Q_0\equiv \sigma_3\otimes \openone_R$ 
where $\sigma_3$ operates in causal space, 
and 
the soft mode action of the associated Goldstone modes $Q=T Q_0 T^{-1}$ 
is precisely the diffusive non-linear sigma model,
here restated for convenience of the reader 
\begin{equation}
S_\sigma 
= \frac{1}{8}\sum_{i,j=1}^D 
\sigma_{ij}^{(0)} 
\int d^3 x
\, 
{\rm Tr}\left(
\partial_i Q \partial_j Q 
\right).
\end{equation}  
In the metallic regime, fluctuations around the saddle point are small. 
At the onset of localization, on the other hand, fluctuations grow uncontrolled and 
rotations start exploring the entire 
field manifold, restoring thus the original symmetry in causal space. 

While the structure of $S_\sigma$ is fixed by general principles, 
details of the matrix degree of freedom depend 
on the system's symmetries. 
So far we have assumed 
the absence of fundamental symmetries. 
In the field theory construction
fundamental symmetry are included by 
``symmetry doubling'' of the matrix dimension. 
The effective description of a system with time reversal symmetry is 
e.g. in terms of a $4R$ dimensional matrix. 
Its entries are then constrained by a symmetry relation  
reminiscent of the symmetry doubling 
(see also Section~\ref{appendix_cft}). 
The same holds for other symmetries, that is,
elements of the ``symmetry doubled'' matrices 
are not independent, and symmetry relations 
fix the field manifold to belong to one of 
ten symmetric spaces~\cite{zirnbauer1996riemannian,altland1997nonstandard}.

\subsection{Sketch of the derivation}
A detailed derivation of the effective action $S_\sigma$
for the dynamical protocol realizing the surface states of the $3d$ quantum spin Hall insulator
can be found in  Sections~\ref{appendix_cft},~\ref{appendix_soft_mode_actions}, 
and~\ref{appendix_derivation_aii_topological_term}, 
and we here only outline the basic 
steps 
(for the corresponding derivation for the quantum Hall simulator 
see e.g. the recent Ref.~\cite{kim2020quantum}). 
Starting out from the 
Floquet operator after gauge transformation,
\begin{align}
\label{usta}
\hat U_\Phi{}
&=
\hat U_{\bold{n}}
\hat U_{\bold{k}},
\end{align}
we focus on the dynamics at long time and length scales.
We assume that 
the action of the local disorder in physical space, 
$\hat U_n$,
in combination with the potential $\hat \Phi$, generated by 
the gauge transformation, induces non-integrability 
in all $1+d_{\rm syn}$ dimensions.  
This assumption is supported by the numerical simulations, as already 
discussed in the previous sections.
To capture the universal long time dynamics, we 
may then erase
system specific details 
introducing an ensemble of 
local spin rotations $U_\bold{n}$
(sharing fundamental symmetries of the system),  
and 
derive a generating functional for the 
\emph{ensemble averaged} correlation function Eq.~(8) in the main text.   

For Floquet systems in class AII the derivation can be organized in terms of   
a color-flavor transformation (cft), whose details 
are exposed in Section~\ref{appendix_cft}. 
Its few line summary is as follows. 
Building on the replica trick and causal doubling, as discussed above,   
we lift 
the Floquet operator from a matrix operating in $2d$ spin space to 
a matrix
$\hat U_\Phi\mapsto \hat U_\Phi\otimes \openone_{2R}$,  
operating in the $2\times 2\times R$ dimensional 
product space 
of spin, causal and replica degrees of freedom. 
The cft then exchanges integrals over 
these 
local  spin rotations, singlet in replica and causal space, for integrals over local rotations  
in replica and causal space, 
and structureless in spin-space.  
The latter conveniently accommodate
the soft modes of the disordered system, viz. 
soft rotations in causal and replica space that are  
singlet in spin-space. 
Formally, the cft is an exact transformation which leads to an alternative 
representation of the generating functional
in matrices that are the local coordinates of 
the matrix degree of freedom $Q(\bold{n})$, discussed above.  
In a final step, see Section~\ref{appendix_soft_mode_actions}, 
the functional is  
expanded in slow fluctuations leading to a soft mode action 
$S=S_\sigma+S_{\rm top}$, 
consisting of the non-linear diffusive $\sigma$ model action  and 
a topological term.

\section{Field theory of class A topological Floquet metal}

Nonlinear sigma models emerge as effective field theories, capturing the low energy sector of disordered single particle systems. 
For static quantum systems, the low energy sector usually describes a narrow window of energies around the Fermi level, 
relevant for the physics at long time and large distance scales. Effective field theories are then 
derived from averaging over disorder ensembles, usually in a Gaussian distribution. 
For time periodic quantum systems, on the other hand, a quasi-energy is defined only modulo $2\pi/T$. Disorder in these systems is usually modeled
by random unitaries drawn from the Haar measure (respectively restrictions of the latter if symmetries are present), and 
there is no distinction between different quasi-energies. 
Zirnbauer~\cite{zirnbauer1996supersymmetry} derived the nonlinear sigma model for time periodic quantum systems
with random onsite unitary disorder using the color flavor transformation, and our discussion below is extensively based on this approach.

Once the effective field theory of a disordered system is known, one may naturally ask whether the system is subject to 
Anderson localization. By now it is well established that Anderson localization can be avoided 
if one of two topological terms, a $\mathbb{Z}_2$ topological term or a WZW term, is present. 
The possibility of the latter is determined 
by the system dimension and symmetries of the nonlinear sigma model target space~\cite{ryu2010topological}. 
Similarly, the existence of topological insulators and superconductors in a given dimension and symmetry class 
can be 
inferred from the presence of robust metallic boundary modes. 
These modes are not subject to Anderson localization and thus define a topological metal. 
In the present work we propose quantum simulators of single copies of topological metals, realized e.g. at 
the {\it isolated} surfaces of topologically non trivial insulators. 

More specifically, $d$-dimensional systems in class A are eligible to topological metallic phases 
if $\pi_{d+1}\left(U(N+M)/U(N)\times U(M)\right)=\mathbb{Z}$. 
Notice here that target manifolds of nonlinear sigma models for Floquet and static systems within the same symmetry class  
are identical, and table 2 of Ryu {\it et al.}~\cite{ryu2010topological} applies for both. 
In our previous work~\cite{kim2020quantum} we derived the nonlinear sigma model action for a $2d$ class A Floquet 
system, 
composed of the conventional diffusive contribution
and a Pruisken $\theta$ term. 
This system is subject to Anderson localization and flows (in the thermodynamic limit) 
to one of the $\mathbb{Z}$ topological insulating phases of the Quantum Hall class. 
Its $1d$ boundary mode is chiral and topologically protected by the winding number of the 
$2d$ insulating bulk. 
The effective theory describing the $1d$ boundary mode can be expressed 
as a product of the $1d$ winding number characterizing the chiral edge mode and 
a $(1+1)d$ WZW term, 
in which the matrix field  of the $1d$ real space coordinate
is deformed (by introduction of an additional homotopy parameter) 
into a trivial field configuration. 

In the present work we derive the effective action for a $3d$ topological metal in class A. 
Its topological term is  
composed of a $3d$ 
winding number multiplied by $(3+1)d$ WZW term, 
in which the matrix field 
parametrized by the $3d$ real space coordinate is extended 
(by introduction of an additional homotopy parameter) to a trivial configuration. 
Since $\pi_4(U(N+M)/U(N)\times U(M))=\mathbb{Z}$, 
the possibility of topologically nontrivial field configurations  
is guaranteed, however, as already discussed constructing a physical model with 
$3d$ winding number is not trivial. We here achieve such model 
employing the idea of synthetic dimensions
engineered via time dependent protocols.
The proposed model can be realized in a quantum walk setting, 
for example, using the optical network shown in Fig.~3 and discussed in the main text. 

An alternative realization of a topological metal is 
via the emergence of a $\mathbb{ Z}_2$ topological term in the 
effective field theory description. 
The latter also protects against Anderson localization, 
and can be realized in one of the real symmetry classes, for example, in a $2d$ class AII system 
for which $\pi_2(O(N+M)/O(N)\times O(M))=\mathbb{Z}_2$. 
In Section D we derive the nonlinear sigma model  for a time reversal symmetric Floquet system 
using the color flavor transformation. We verify that the $\mathbb{Z}_2$ topological term is 
written as the product of two $(2+1)d$ WZW terms of matrix fields, one involving the translational invariant 
part of the time evolution operator and the other the sigma model field degree of freedom. 
In combination with the $\mathbb{Z}$ example,
this completes our derivation of low energy  effective field theories for 
topological Floquet metals, their model realizations, 
and numerical confirmation of the absence of Anderson localization. 

In the remaining part of this Section we discuss the field theories for class A and class AII systems. 
We introduce the color flavor transformation for class A, and the derivation of a $(3+1)d$ WZW term is detailed 
in Section C. In Section D we present the color flavor transformation for class AII systems and the derivation of a 
$\mathbb{Z}_2$ topological term, and its alternative representation in terms of a WZW term.

\subsection{Color-flavor transformation class A Floquet systems} 
Consider a Floquet system described by the single time step evolution operator $\hat U_0$, to which 
onsite disorder is introduced by adding  to the time evolution locally uncorrelated random phases.
The microscopic action is then of the following general form, $Z = \int D(\psi,\bar \psi) \exp (-S)$, 
with
\begin{align}
S &= \int d^d\bold x \, \bar \psi_{\bold x}^{+a} (\hat G_{+}^{-1})_{\bold {xx'}}\psi_{\bold x'}^{+a} + \bar \psi_{\bold x}^{-a} (\hat G_{-}^{-1})_{\bold {xx'}} \psi_{\bold x'} ^{-a},\label{app_PsiUFunctional} \\
&(\hat G_{+}^{-1})_{\bold {xx'}} = \delta_{\bold{xx'}}-e^{i\phi_{\bold x}}(\hat U_0)_{\bold {xx'}}, \\
&(\hat G_{-}^{-1})_{\bold {xx'}} = \delta_{\bold{xx'}}-(\hat U_0^\dg ) _{\bold {xx'}}e^{-i\phi_{\bold x'}}, 
\end{align}
where $\exp(i\phi_{\bold x})$ are the uniformly distributed phases, and 
`$a$' carries spin, replica, and particle-hole indices. 
Upon introduction of adequate source terms (see e.g. Ref.~\cite{kim2020quantum} for details), 
Eq.~\eqref{app_PsiUFunctional} 
 allows for the convenient generation of 
products 
 $\langle G^+ G^-\rangle$, 
 e.g. required for the calculation of Eq.~(8) in the main text.
 The disorder averaging of the generating function $\langle Z \rangle = \prod_{\bold x} \frac{1}{2\pi} \int d\phi_{\bold x} Z(\{ \phi_{\bold x}\})$ is performed by the color flavor transformation: 
\begin{align}
& \frac{1}{2\pi} \int d\phi_{\bold x} e^{ \bar \psi_{\bold x}^{+a} e^{i\phi_{\bold x}}\varphi_{\bold x}^{+a} + \bar \varphi_{\bold x}^{-a} e^{-i\phi_{\bold x}} \psi_{\bold x} ^{-a} }  \\
&=  \int D(Z_{\bold x},\tilde Z_{\bold x}) e^{-\tr\ln (1- \tilde Z_{\bold x} Z_{\bold x} )} e^{ \bar \psi_{\bold x}^{+a} Z_{\bold x} ^{ab} \psi_{\bold x} ^{-b} + \bar \varphi_{\bold x} ^{-a} \tilde Z_{\bold x}^{ab} \varphi_{\bold x} ^{+b} }, \label{cft_identity}
\end{align}
applying for every position $\bold x$. 
Here the Floquet operator $\hat U_0$ has been absorbed into the newly defined Grassmann fields 
$\varphi_{\bold x}^{+a}=(\hat U_0 \psi)_{\bold x}^{+a}$ 
and 
$\bar \varphi_{\bold x}^{-a}=(\bar \psi \hat    U_0^\dagger)_{\bold x}^{-a}$, 
and   
the domain of integration is defined by the condition $\tilde Z = - Z^\dg$. 
The matrix field $Z_{\bold x}^{ab}$ connects retarded (+) and advanced (-) Grassmann field, 
indicating that its spatial fluctuation encodes diffusion in the long distance limit. 
Its role becomes more clear by introducing $Q = T \tau_3 T^{-1}$ with 
\begin{align*}
T \equiv \begin{pmatrix}
\openone & Z \\ \tilde Z & \openone 
\end{pmatrix}_{\rm RA},
\end{align*} 
and anticipating that the effective field theory 
is expressed in terms of the matrix field $Q$ with nonlinear constraint $Q^2 =\openone$. 
More specifically, $Z$ defines linear coordinates on the symmetric space $U(2R)/U(R)\times U(R)$ 
with $Z=0$ representing the `north pole', $Q=\tau_3$, and $Z \rightarrow \infty$ the `south pole', $Q = -\tau_3$. 
After integration over Grassman fields, we obtain the  class A action entirely expressed in 
terms of the linear coordinates,
\begin{align*}
S[Z,\tilde Z] = -\tr \ln (1-\tilde Z Z) + \tr \ln (1-\tilde Z U Z U^\dg). 
\end{align*}
In Ref.\cite{kim2020quantum} we derived the low energy effective action of a $2d$ Floquet topological insulator, 
composed of the conventional diffusive term accompanied by a Pruisken $\theta$ term. This allowed us to  
confirm that the maximally disordered Floquet system belongs to the integer quantum Hall universality class. 
In the following sections, we derive the effective theory for a $3d$ Floquet system. This turns out to 
contain a Chern-Simons action, and thus is within the same universality class 
as Weyl semimetals without intervalley scattering.

\subsection{Soft-mode actions for class A}

\label{appendix_3d_classA}

A straightforward manipulation of block matrices brings the above representation of the action into the form ~\cite{kim2020quantum}
\beq
S &=& \frac{1}{2}\sum_{s=\pm} \mathrm{Tr}\, \mathrm{ln} (1+ X_s P^s), \\
X_- &\equiv & \hat T^{-1} [\hat U_0,\hat T] \hat U_0^\dg, \,\,\,\,\, X_+ \equiv \hat T^{-1} [\hat U_0^\dg,\hat T] \hat U_0,
\eeq
where $P^\pm = (\openone \pm \tau_3)/2$ are projectors onto retarded and 
advanced sectors of the theory. 
To simplify notation, we will  in the following drop the index `$0$' of the Floquet operator $\hat U_0$. 
An expansion of the log-function to the third order is necessary, 
\beq
S &=& S^{(1)} + S^{(2)} + S^{(3)}, 
\eeq 
where
\beq
S^{(1)} &=&  \frac{1}{2}\sum_{s=\pm} \mathrm{Tr}(X_sP^s),\nonumber \\ S^{(2)} 
&=&  -\frac{1}{4}\sum_{s=\pm} \mathrm{Tr}((X_sP^s)^2),\nonumber \\  
S^{(3)} &=&  \frac{1}{6}\sum_{s=\pm} \mathrm{Tr}((X_sP^s)^3). \nonumber 
\eeq
In the following sections, using the Wigner transformation the continuum representation of the action is obtained. 

\subsubsection{The first order terms $S^{(1)}$}

Consider the `-' contribution to the action $S^{(1)}$,
\begin{align*}
&S^{(1)-} 
\nonumber\\
&= \frac{1}{2}\mathrm{Tr}\left( (\hat T^{-1} \hat U\hat T\hat U ^\dg - \mathbb{I}) P^-\right), 
\nonumber\\
&= \frac{1}{2}\mathrm{Tr}\left( \int_0^t ds ( \hat T^{-1} \partial_s \hat U\hat  T\hat U ^\dg + \hat T^{-1} \hat U\hat T \partial_s \hat U ^\dg)P^-\right), 
\nonumber\\
&= \frac{1}{2}\int_0^t ds\, \mathrm{Tr}\left( \left( [ \hat T^{-1}, \partial_s \hat U \hat U^\dg ] \hat U\hat T\hat U^\dg \right) P^-\right), 
\nonumber\\
&
= -\frac{1}{2}\int_0^t ds\, \mathrm{Tr}\left( \left( [ \hat T^{-1}, \hat\psi_s^-] [\hat U,\hat T]\hat U^\dg + [ \hat T^{-1}, \hat\psi_s^-] \hat T \right) P^-\right), 
\end{align*}
where in the third equality  $ \hat U \partial_s \hat U^\dg = -\partial_s \hat U \hat U^\dg \equiv \psi_s^- $ is used. Note that $\hat U$ has no structure in 
replica space, thus $[\partial_s \hat U^\dg, P^-] = 0$ is used. The commutator in the third line can be Moyal expanded up to the third derivatives, 
\beq
[ \hat T^{-1}, \hat\psi_s^-] [\hat U,\hat T]\hat U^\dg  &\simeq & (\partial_i T^{-1} \partial_i \psi_s^-)  (\partial_j U \partial_j T U^\dg) ,\nonumber 
\eeq
and
\beq
\left[ \hat T^{-1}, \hat\psi_s^-\right] \hat T  &\simeq &
  i(\partial_i T^{-1} \partial_i  \psi_s^-)T - \frac{i}{24} (\partial^3_{ijk} \psi_s^- \partial^3_{ijk} T^{-1}) T, \nonumber 
\eeq
where distinction between an operator and a function is made by hat on symbols.  $\partial_i U \equiv \partial_{k_i} U$ and $\partial_i T \equiv \partial_{x_i} T$ and sum over index $i,j,k$ is implicitly assumed. By plugging in the above into the action, 
\begin{multline*}
S^{(1)-} = -\frac{i}{2}\int_0^t ds \,\mathrm{Tr}\left(( \partial_i T^{-1}   T ) (\partial_i \psi_s^-)   P^-\right) \\ +
 \frac{1}{2} \int_0^t ds\, \mathrm{Tr} \left((\partial_i T^{-1}\partial_{j} T ) (\partial_i \psi_s^- \psi_j^-)P^- \right) \nonumber \\
 +
\frac{i}{48}\int_0^t ds \, \mathrm{Tr}\left( ( \partial^3_{ijk} T^{-1}  T) (\partial^3_{ijk} \psi_s^- ) P^- \right),  
\end{multline*}
where the first and third term vanish after the momentum integration because $\psi_s^-$ a is a periodic function in momentum. The second term, which is obtained using $\partial_j U U^\dg = - U \partial_j U^\dg=-\psi_j^-$, is,
\beq
S^{(1)-}_{\mathrm{2nd}} &=& \frac{1}{2} \int_0^t ds\, \mathrm{Tr} (\partial_i T^{-1} \partial_j T P^- ) \mathrm{tr} (\partial_i \psi_s^- \psi_j^-).\nonumber \\ \label{S12}
\eeq
Eq.~\eqref{S12} is identical to the one in \cite{kim2020quantum}, from which the Pruisken action and a part of diffusive action is derived.

\subsubsection{The second order terms $S^{(2)}$}

Using $X_- = \hat T^{-1} [\hat U,\hat T]\hat U^\dg  = \hat T^{-1} \hat U [\hat T,\hat U^\dg] \rightarrow i (T^{-1}* U )* \partial_i T \partial_i U^\dg$, and Moyal expand to the third derivatives,
\begin{multline*}
(T^{-1}* U )* \partial_i T \partial_i U^\dg \\ \simeq T^{-1} U \partial_i T \partial_i U^\dg  + \frac{i}{2}(\partial_k T^{-1} \partial_k U ) \partial_i T \partial_i U^\dg \\ + \frac{i}{2}(\partial_k T^{-1} U ) \partial_i T \partial^2_{ik} U^\dg  - \frac{i}{2}(T^{-1} \partial_k U ) \partial^2_{ik} T \partial_{i} U^\dg , 
\end{multline*}
which is then plug in to $S^{(2)-}$, 
\begin{align}
& S^{(2)-} = \frac{1}{4} \mathrm{Tr} \left((T^{-1}\partial_i T)P^- (T^{-1}\partial_j T)P^-  (U\partial_i U^\dg U\partial_j U^\dg) \right) \nonumber \\  &+
 \frac{i\epsilon^{ijk}}{4} \mathrm{Tr} \left( (\partial_kT^{-1}\partial_i T )P^- (T^{-1}\partial_j T)P^-( \partial_k U \partial_i U^\dg U\partial_j U^\dg)\right) \nonumber \\
 &+ \frac{i\epsilon^{ijk}}{4}\mathrm{Tr} \left( (\partial_kT^{-1}\partial_i T)P^- (T^{-1}\partial_j T)P^- ( U\partial^2_{ik}U^\dg U\partial_j U^\dg) \right) \nonumber \\  &-\frac{i\epsilon^{ijk}}{4}\mathrm{Tdr} \left((T^{-1}  \partial^2_{ik} T )P^- (T^{-1}\partial_j T)P^- (\partial_k U\partial_i U^\dg U \partial_j U^\dg) \right).\label{S23b} 
\end{align}
 The Levi-Civita symbol is introduced because only antisymmetric combinations are nonzero for the 3-dim TFM unitary operator. Thus, the third and fourth term in \eqref{S23b} is zero. Introducing $A_i=T^{-1}\partial_i T$, 
\begin{align*}
S^{(2)-}
&= \frac{1}{4}\mathrm{Tr} \left( (A_i P^- A_j P^-)(\psi^-_i \psi^-_j) \right) \nonumber \\ 
&+  \frac{i}{4} \epsilon^{ijk}\mathrm{Tr}\left((A_k A_i P^- A_j P^- ) (\psi_k^- \psi^-_i \psi^-_j)\right),
\end{align*}
Note that $\epsilon^{ijk}A_j A_k = -\epsilon^{ijk}\partial_j A_k$.  On the other hand, 
\begin{align*}
S^{(2)+}
&= \frac{1}{4}\mathrm{Tr} \left( (A_i P^+ A_j P^+)(\psi^+_i \psi^+_j) \right) \nonumber \\ 
&+  \frac{i}{4} \epsilon^{ijk}\mathrm{Tr}\left((A_k A_i P^+ A_j P^+ ) (\psi_k^+ \psi^+_i \psi^+_j)\right).
\end{align*}
Employing that 
$\mathrm{tr}(\psi_i^+\psi_j^+) = \mathrm{tr}(\psi_i^-\psi_j^- )$ 
and $\mathrm{tr}(\psi_i^+\psi_j^+\psi_k^+) = - \mathrm{tr}(\psi_i^-\psi_j^- \psi_k^-)$, 
the latter two terms of $S^{(2)\pm}$ can be combined as
\begin{align}
\label{eq:S_2_2nd}
S^{(2)} &= \frac{1}{4}\sum_{s=\pm} \mathrm{Tr} (A_iP^s A_j P^j) \\
&-\frac{i\epsilon^{ijk}}{4}\sum_{s=\pm} s\,\mathrm{Tr}\left( \partial_k A_i P^s A_j P^s\right) \mathrm{tr}(\psi_i^+\psi_j^+\psi_k^+), \nonumber
\end{align}
where the first term becomes part of the diffusive action 
(see Eq.(30) of Ref.~\cite{kim2020quantum}), 
and the second term constitutes the first part 
of the Chern-Simons action.

\subsubsection{The third order terms $S^{(3)}$}

The third order expansion of the log yields, 
\begin{align*}
S^{(3)-} &= \frac{i^3}{6} \epsilon^{ijk}\mathrm{Tr}\left( (T^{-1}U\partial_i T\partial_iU^\dg)P^- \right. \nonumber \\
&\times \left. (T^{-1}U\partial_j T\partial_jU^\dg)P^- (T^{-1}U\partial_k T\partial_k U^\dg)P^-\right), \\
&=-\frac{i}{6} \epsilon^{ijk}\mathrm{Tr}(A_i P^- A_j P^- A_k P^- )\,\mathrm{tr} (\psi^-_i \psi^-_j \psi^-_k ), 
\end{align*}
and similarly, 
\begin{align*}
S^{(3)+} &=-\frac{i}{6} \epsilon^{ijk}\mathrm{Tr}(A_i P^+ A_j P^+ A_k P^+ )\,\mathrm{tr} (\psi^+_i \psi^+_j \psi^+_k ).
\end{align*}
Combining the latter with the 2nd piece in Eq.~(\ref{eq:S_2_2nd}), we arrive at 
\begin{multline*}
S^{(3)} = -\frac{i\epsilon^{ijk}}{6}\sum_{s=\pm} s\, \mathrm{Tr}(A_i P^s A_j P^s A_k P^s )\,\mathrm{tr} (\psi^+_i \psi^+_j \psi^+_k ),
\end{multline*}
which constitutes the second part of the Chern-Simons action. 
Introducing the winding number 
$\nu_3 = \frac{1}{24\pi^2}\int d^3 k \,\epsilon^{\mu\nu\rho}\mathrm{tr}(\psi^+_\mu \psi^+_\nu\psi^+_\rho)$, 
the Chern-Simons action can be expressed in the following form: 
\begin{align*}
S_{\rm top}
&= \nu_3 \int_{\partial B_4} \left( w_{\rm CS} [ AP^+] - w_{\rm CS}[AP^-] \right), 
\end{align*}
where $w_{\rm CS}[A] = \frac{i}{8\pi}\left(A\wedge dA + \frac{2}{3} A\wedge A\wedge A \right)$. As a result we obtain the 
topological action summarized in Eqs.~(23)
and (25) in the main text. 
Notice that the presence of the Chern-Simons action can be anticipated 
for a $3d$ class A system, noting that the homotopy group $\pi_{3+1} (U(2R)/U(R) \times U(R) )=\mathbb{Z}$ is nontrivial.
That is, the field $T=T(\bold x)$ can be extended 
from $3d$ space to the matrix field $Q=Q(x_0,\bold x)$ in one dimension higher, and  
the topological action can then be expressed as a WZW term 
(see Ref.~\cite{zhao2015general}): 
\begin{align*}
S_{\rm top}&= \frac{i\nu_3}{128\pi} \int_{B_4} {\rm Tr}\, \left[ Q(\wedge dQ)^4 \right]. 
\end{align*}

\section{Field theory of class AII topological Floquet metal}
In this Section we provide details on the effective field theory for class AII Floquet systems. 
We first present the color flavor transformation for the latter, 
from which the disorder averaged effective action is derived in D.1. 
The softmode manifold of class AII systems is specified in D.2, and in sections D.3 and D.4 
we present the nonlinear sigma model 
composed of a diffusive and a topological term. 
The latter is expressed as the product of two WZW terms,  
measuring the topological content of mappings 
from momentum and real space, respectively. 
We provide the connection of the topological term and the color flavor action in D.5, 
completing thus the derivation of effective field theories for Floquet topological metals. 
Conventions used throughout the sections are as follows.
We use symbols $\sigma_j$ to indicate sublattice space, 
$\tau_j$ for retarded-advanced space, and $s_j$ for particle-hole space. 
Vectors $(x_0,\bold x) = (x_0, x_1,\cdots, x_d)$ summarize the spatial coordinate and homotopy parameter, 
and correspondingly,
$(k_0,\bold k) = (k_0, k_1,\cdots, k_d)$, for momentum coordinates including a homotopy parameter.

\subsection{Color-flavor transformation for class AII Floquet systems}
\label{appendix_cft}
We recall that the $2\times 2$ dimensional time evolution operator in class ${\rm AII}$ satisfies 
the time-reversal symmetry constraint $\sigma_2 \,{\cal U_F}^T \sigma_2 = {\cal U_F}$.
It can be written as a product
\begin{equation}
{\cal U_F} = V U_0 \bar V, 
\quad 
\bar V = \sigma_2 V^T \sigma_2, 
\quad 
\bar U_0 = \sigma_2 U_0^T \sigma_2 \equiv U_0,
\end{equation}
where $U_0$ is a non-random part while $V$ and $\bar V$ encode the unitary disorder. For the simplicity of notation the index zero in the clean part of the Floquet operator $U_0$ will be dropped in the following discussion. 
Concentrating e.g. on the contribution of the retarded (+) sector 
to the action~\eqref{app_PsiUFunctional}, the above decomposition of ${\cal U_F}$ 
allows to rewrite
\begin{equation}
S_+ 
= 
\bar\psi_{+1} \,\psi_{+1} + \bar\psi_{+2} \, \psi_{+2} 
- 
\bar\psi_{+1} e^{i\phi_+} V \psi_{+2} - \bar\psi_{+2} U \bar V \psi_{+1},
\end{equation} 
where we introduced the two-component spinor 
$\bar\psi_+=(\bar\psi_{+1}, \bar\psi_{+2})$ and similar 
for $\psi_+$. A constant source term $\phi_+$ is introduced. 
We then arrange contributions containing $\bar V$ 
using time reversal symmetry, 
\begin{align}
\bar\psi_{+2} U \bar V \psi_{+1} 
&
= (\bar\psi_{+2} U \sigma_2 V^T \sigma_2 \psi_{+1})^T 
\nonumber\\
&= 
-\psi_{+1}^T  \sigma_2 V \sigma_2 U^T \bar V \bar\psi_{+2}^T 
\nonumber\\
&= 
-(\psi_{+1}^T  \sigma_2)\, V\, (U \sigma_2 \bar\psi_{+2}^T), 
\end{align} 
and upon introducing spinors
\begin{equation}
\varphi_+^T = \left(\bar\psi_{+1}, i\, \psi_{+1}^T  \sigma_2 \right), 
\quad
\chi_+ = \left(\begin{array}{c}
e^{i\phi_+} \psi_{+2} \\  U i \sigma_2 \bar\psi_{+2}^T
\end{array}\right),
\end{equation}
we can cast this action into the simpler form 
\begin{equation}
S_+ =  \bar\psi_{+1} \, \psi_{+1} + \bar\psi_{+2} \, \psi_{+2} - \varphi_+^T V \chi_+.
\end{equation}
Proceeding along the same lines for the contribution of the advanced 
 sector to the action~\eqref{app_PsiUFunctional},
\begin{equation}
S_- = \bar\psi_{-1} \,\psi_{-1} + \bar\psi_{-2} \, \psi_{-2} 
- \bar\psi_{-1} \bar V^\dagger U^\dagger \psi_{-2} 
- \bar\psi_{-2} V^\dagger e^{-i\phi_-} \psi_{-1},
\end{equation} 
the disorder term $\propto V^\dagger$ can be reorganized 
employing time reversal symmetry,
\begin{align}
&\bar\psi_{-1} \bar V^\dagger U^\dagger \psi_{-2} 
\nonumber\\
&= (\bar\psi_{-1} \sigma_2 V^*\sigma_2 U^\dagger \psi_{-2})^T 
\nonumber\\
&=
-(\psi_{-2}^T U^* \sigma_2) V^\dagger (\sigma_2 \bar\psi_{-1}) 
\nonumber\\
&= 
-(\psi_{-2}^T \sigma_2 U^\dagger) V^\dagger (\sigma_2 \bar\psi_{-1})
\end{align}
and upon 
introducing the spinors
\begin{equation}
\chi_-^T = (\bar\psi_{-2}e^{-i\phi_-},  i \psi_{-2}^T \sigma_2 U^\dagger),
\quad
\varphi_- = \left(\begin{array}{c}
\psi_{-1} \\  i \sigma_2 \bar\psi_{-1}^T,
\end{array}\right),
\end{equation}
into the form
\begin{equation}
S_- 
=  \bar\psi_{-1} \, \psi_{-1} + \bar\psi_{-2} \, \psi_{-2} - \chi_-^T V^\dagger \varphi_-.
\end{equation}

With these preparations we can now apply the color-flavor transformation 
to the action $S=S_+ + S_-$ by averaging over $V$ using \eqref{cft_identity}. 
This amounts to exchanging matrices $V$, $V^\dagger$, with structure in spin space, 
for matrices $Z$, $\tilde Z$, with structure in replica and causal space. 
That is, 
\begin{align}
S_1 
&= \bar\psi_{+1} \, \psi_{+1} +  \bar\psi_{-1} \, \psi_{-1} - \varphi_+^T Z \varphi_-, 
\nonumber\\
S_2 
&=\bar\psi_{+2} \, \psi_{+2} +  \bar\psi_{-2} \, \psi_{-2} - \chi_-^T \tilde Z \chi_+, 
\end{align}
and we notice that $\psi_{\pm 1}$ can be expressed 
entirely in terms of $\varphi_\pm$,
 and correspondingly, $\psi_{\pm 2}$  
entirely in terms of $\chi_\pm$. 
The subsequent path integral can therefore 
be split into two independent integrals over $\varphi$ and $\chi$ 
with actions $S_1$ and $S_2$, respectively.

To proceed, 
we organize $\pm$ indices of vectors $\varphi$ and $\chi$ into a two dimensional space, 
in the following referred to as `particle-hole (ph)' space, and  
introduce the matrix
$s_1 \equiv \sigma_1^{\rm ph}$, 
operating in this space,
with 
$s_1^T s_1 = 1$. 
We then notice that 
\begin{align}
\varphi_+^T\, (i\sigma_2 \otimes s_1)\, \varphi_+ 
&= 2 \bar \psi_{+1} \psi_{+1}, 
\nonumber\\
\varphi_-^T\, (i\sigma_2 \otimes s_1)\, \varphi_- 
&= 2 \bar \psi_{-1} \psi_{-1}, 
\nonumber\\
\chi_+^T \, e^{-i\phi_+} (i\sigma_2 \otimes s_1) U^\dagger \, \chi_+ 
&= 2 \bar \psi_{+2} \psi_{+2}, 
\nonumber\\
\chi_-^T \, e^{i\phi_-} (i\sigma_2 \otimes s_1) U^T \, \chi_- 
&= 2 \bar \psi_{-2} \psi_{-2},
\end{align}  
which allows us 
to express actions $S_{1,2}$ in the form 
\begin{equation}
\label{eq:S_12_AII}
S_1 
= \frac 12 (\varphi_+^T, \varphi_-^T) \left(\begin{array}{cc}
(i\sigma_2 \otimes s_1) & - Z \\  Z^T  & (i\sigma_2 \otimes s_1)
\end{array}\right) 
\left(\begin{array}{c}
\varphi_+ \\ \varphi_-
\end{array}\right), 
\end{equation}
and
\begin{align*}
S_2 
&=
\frac 12 (\chi_+^T, \chi_-^T) \\
&\times
\left(\begin{array}{cc}
e^{-i\phi_+}(i\sigma_2 \otimes s_1) U^\dagger & \tilde Z^T  \\ -\tilde Z  & e^{i\phi_-} (i\sigma_2 \otimes s_1) U^T \
\end{array}\right) 
\left(\begin{array}{c}
\chi_+ \\ \chi_-
\end{array}\right),
\end{align*} 
were we used that  
$\varphi_+^T Z \varphi_- = -\varphi_-^T Z^T \varphi_+$ 
and similar for $\chi$.
In a final step we 
  then complete Gaussian integrals 
 over $\varphi$ and $\chi$. 
Accounting for a nontrivial Jacobian $J(\phi_+,\phi_-) = ( e^{-i(\phi_+ - \phi_-)})^{-1}$ 
resulting from the change of integration variables,  
we arrive at
\begin{align}
{\cal Z} 
& \propto  
J (\phi_+,\phi_-)\, {\rm det}^{1/2}\left(\begin{array}{cc}
i \sigma_2 \otimes s_1 & - Z \\  Z^T  & i\sigma_2  \otimes s_1
\end{array}\right)
\times\nonumber\\
&\times 
{\rm det}^{1/2}\left(\begin{array}{cc}
e^{-i\phi_+}(i \sigma_2  \otimes s_1) U^\dagger &  \tilde Z^T  
\\ 
- \tilde Z & e^{i\phi_-} (i\sigma_2  \otimes s_1)U^T \
\end{array}\right).
\end{align}
Finally, restoring the measure
$\mu(Z,\tilde Z) = {\rm det}(1-\tilde Z Z)^{-1}$ 
of the color-flavor transformation, 
we arrive at 
${\cal Z}=e^{S}$, with
\begin{align}
\label{app_eq:S_ZZ}
S
&= 
-{\rm tr} \ln (1-\tilde Z Z) 
 +
\frac 1 2 {\rm tr}\ln( 1 - Z^T s_1 Z s_1^T )  
\nonumber\\
&\quad
+
\frac 1 2 {\rm tr}\ln( 1 - \tilde Z U s_1^T e^{i\phi_+} 
\tilde Z^T e^{- i\phi_-} s_1 U^\dagger ). 
\end{align} 
Upon color-flavor transformation (enabling the disorder averaging), 
the original action for `microscopic' Grassmann fields is expressed in terms of 
the matrix fields $Z$, $\tilde Z$. These   
are more convenient to extract the low energy content of the theory, 
and to access the low energies we 
subject the action to a saddle point analysis.  
Allowing for  
(soft) spatial fluctuations around the saddle points we then derive 
the low energy effective action. We notice that the target space of the matrix field is 
topologically non trivial and  the generating function must include the sum over topologically distinct configurations, 
weightening topologically different sectors by a topological contribution to the action. 
We present the effective action in Sections~\ref{appendix_soft_mode_actions} and~\ref{appendix_z_2 from_wess_zumino_witten}, 
and the derivation of the topological term (starting out from Eq.\eqref{app_eq:S_ZZ}) 
in Sec.\ref{appendix_derivation_aii_topological_term}.

\subsection{Soft-mode manifold for class AII}

We then notice that 
in the zero frequency limit, $\phi_\pm \to 0$, 
 action Eq.~(\ref{app_eq:S_ZZ}) 
is minimized by homogeneous configurations 
$Z,\tilde Z$ related via
\begin{equation}
\label{app_eq:Z_soft}
\tilde Z = s_1^T Z^T s_1. 
\end{equation}
The soft mode action reads
\begin{equation}
\label{app_eq:ZZ_classAII}
S[Z,\tilde Z] 
= - \frac 12 {\rm tr} \ln (1-\tilde Z Z) + 
\frac 12 {\rm tr}\ln( 1 - \tilde Z U   Z U^\dagger ),
\end{equation}
which apart from the factor $1/2$ and symmetry constraint Eq.~\eqref{app_eq:Z_soft},
 is the same as that encountered for class ${\rm A}$ systems. 
 (See e.g. Ref.~\cite{kim2020quantum} for a discussion in the context of the above mentioned AFAI.)

To further elaborate on the soft mode manifold, we introduce
matrices, $\tau_3  \equiv \sigma_3^{\rm RA}$, 
\begin{equation}
Q = T \tau_3 T^{-1}, 
\qquad 
T = \left(\begin{array}{cc}
1 & Z \\ \tilde Z & 1
\end{array}\right)_{\rm RA},
\end{equation} 
satisfying the symmetry constraint~\cite{Efetov-book}
\begin{align}
&\bar Q = Q, \quad \bar Q \equiv C^T Q^T C, 
\quad 
C = \tau_3 \otimes s_1. 
\end{align}
The explicit expression for the $Q$-matrix read,
\begin{equation}
Q = 
\left(\begin{array}{cc}
\displaystyle \frac{1+ Z \tilde Z }{1- Z \tilde Z } & \displaystyle -2Z \frac{1}{1- \tilde Z Z } \\ 
\displaystyle 2\tilde Z \frac{1}{1- Z \tilde Z } & \displaystyle -\frac{1+ \tilde  Z Z }{1- \tilde Z Z}
\end{array}\right)_{\rm RA}, 
\end{equation}
where we recall the relation $A ({1- B A })^{-1} =  ({1-A B  })^{-1}A $. 
One can now verify that owing to~(\ref{app_eq:Z_soft})  
the relation $C^T Q^T C = Q$ holds. 
This defines  the soft mode manifold of class ${\rm AII}$ systems.

\subsection{Soft-mode actions for class AII}
\label{appendix_soft_mode_actions}

Following the steps discussed in Ref.~\cite{kim2020quantum}, we can apply a gradient expansion to 
derive a diffusive $\sigma$ model action for $Q$ fields. The non trivial properties 
of the dynamical protocols for the simulators of topological surface states 
are encoded in an additional topological contribution 
to the soft mode action. Their derivation is discussed in more detail below. 
The soft mode action 
for $Q$-matrix fields satisfying AII symmetry constraint $Q=\bar Q$ 
therefore reads
$S=S_\sigma+S_{\rm top}$,
 where
\begin{align}
\label{app_eq:S_AII}
S_\sigma[Q] 
&= - \frac{1}{8} \sum_{i,j=1}^2 \sigma_{ij}^{(0)}\int d^2 x\, {\rm tr} (\partial_i Q \partial_j Q), 
\\
\label{app_eq:S_AII_top_term}
S_{\rm top}[Q] 
&= \frac{ i \theta}{\pi} \, \Gamma[g]\Bigl|_{g(x_0=0,{\bf x})=Q({\bf x})},
\end{align}
with conductivity tensor $\sigma_{ij}^{(0)}
 =\frac{1}{2}
 \int d^2\bold k\, \tr(\partial_{k_i} U_\bold{k}\partial_{k_j} U_\bold{k}^{-1})$ and 
topological angle, $\theta = 0,\pi$ defined in Eq.~(\ref{eq:theta}).
$\Gamma[g]$ in Eq.~\eqref{app_eq:S_AII_top_term} is a WZW functional  
\begin{align}
\label{eq:Gamma_+}
&\Gamma[g] 
= \frac{1}{24\pi} \int_0^1 dx_0 \int d^2 \bold x \epsilon^{\mu\nu\rho} 
{\rm tr}\left( g^{-1}\partial_\mu g g^{-1}\partial_\nu g g^{-1}\partial_\rho g \right), 
\end{align}
here defined for matrices $g$ from the orthogonal group ${\rm O}(4R)$, 
satisfying the symmetry relation $\bar g = C^T g^T C = g^{-1}$, 
and with constraint 
$g(x_0=1,{\bf x})= \tau_3$. 

Several comments are here of order.  
(i) We remind that the WZW action depends only on the boundary 
value of the group field, $g(x_0=0,{\bf x})$. 
Given that the homotopy group $\pi_2({\rm O}(4R))=0$ is trivial, 
an extension to the third dimension $x_0$ with constant boundary value  
at $x_0=1$ is always possible.  
(ii) As we show below, $\Gamma[g]$ is only defined modulo $\pi$.
However, recalling that $\theta=0,\pi$, the exponentiated action   
$e^{-S_{\rm top}[Q]} = \pm 1$ is well defined. 
(iii) Finally, as already discussed in the main text,  
the possibility to add the topological term 
$S_{\rm top}[Q]$ to the diffusive action 
is related to the nontrivial homotopy group of 
the coset space, 
$\pi_2({\rm O}(4R)/{\rm O}(2R) \times {\rm O}(2R)) 
= \mathbb{Z}_2$.  

We conclude our discussion, showing 
that $S_{\rm top}[Q]$ is indeed a $\mathbb{Z}_2$ topological term. 
That is, it (a) does not change under
local variations of $Q$ 
which stay in the same equivalence class, 
and (b) takes values $0$ or $i\pi$.
Starting out with (a), we first notice that 
 variation of the WZW action with respect to arbitrary 
 fluctuations $g$ gives
\begin{equation}
\delta \Gamma[g] 
= \frac{1}{8\pi} 
\int d^2 \bold x \epsilon^{\mu\nu} 
{\rm tr}( g^{-1}\delta g g^{-1}\partial_\mu g g^{-1}\partial_\nu g)\Bigl|_{x_0=0}.
\end{equation}
Recalling then that on 
the coset manifold $Q^{-1} = Q$, 
we can use $Q \delta Q = -\delta Q Q$ 
and $Q \partial_\mu Q = - \partial_\mu Q Q$,
to find 
\begin{align}
\epsilon^{\mu\nu} 
&
{\rm tr}( Q\delta Q\,\, Q\partial_\mu Q\,\, Q\partial_\nu Q) 
\nonumber\\
&= 
- \epsilon^{\mu\nu} {\rm tr}( \delta Q Q \,\, \partial_\mu Q Q \,\,\partial_\nu Q Q) 
\nonumber\\
&= 
- \epsilon^{\mu\nu} {\rm tr}( Q \delta Q \,\, Q  \partial_\mu Q \,\, Q\partial_\nu Q )
=0.
\end{align}
That is, $\delta S_{\rm top}[Q] = 0$ for any local variations of the $Q$-matrix. 
For (b), we show that $\Gamma[Q]=0,\pi$ modulo contributions $2\pi$.  
To this end, we extend $g(x_0>0,{\bf x})$ to $x_0 \in [-1,0)$ defining
 $g(x_0<0, {\bf x}) = \bar g (x_0>0,{\bf x})$ which, recalling that $g|_{x_0=0} = Q = \bar Q$, 
 is continuous at $x_0=0$. 
Denoting then by $\Gamma_+[g]$ and $\Gamma_-[g]$ the two 
WZW terms 
in Eq.~(\ref{eq:Gamma_+}) for which integration over $u$ 
is carried out over positive and negative values, respectively,  
(b) follows from the following two observations.  
First, $\Gamma_+[g]+\Gamma_-[g]=2\pi\mathbb{Z}$ 
is counting the winding number of the orthogonal group  
(which reflects the non-trivial homotopy $\pi_3({\rm O}(4n)) = \mathbb{Z}$), and 
second, $\Gamma_+[g]=\Gamma_-[g]$. 
To show the latter, we notice that 
for $x_0>0$ we can write $g'(-x_0) = - \partial_{x_0} g(-x_0) = - \bar g'(x_0)$, 
and therefore 
\begin{align}
\label{eq:Gamma_-}
&\Gamma_-[g] 
\nonumber\\
&= \frac{1}{24\pi} \int_{-1}^0 dx_0 \int d^2 \bold x \epsilon^{\mu\nu\rho} 
{\rm tr}\left( g^{-1}\partial_\mu g g^{-1}\partial_\nu g g^{-1}\partial_\rho g \right)  
\nonumber\\
&=
 - \frac{1}{24\pi} \int_{0}^1 dx_0 \int d^2 \bold x \epsilon^{\mu\nu\rho} 
{\rm tr}\left( \bar g^{-1}\partial_\mu \bar g \bar g^{-1}\partial_\nu \bar g \bar g^{-1}\partial_\rho \bar g \right) 
\nonumber\\
&=
- \,\Gamma_+[g^{-1}] = \Gamma_+[g],
\end{align}
where 
in the second line we set $x_0\to -x_0$, and in the third line 
$\bar g = g^{-1}$. The value $\Gamma[g]=0,\pi$ 
can be thought as the $\mathbb{Z}_2$ topological index of a matrix $Q=g(x_0=0,\bold x)$, 
and is a direct analog of the winding number defined 
by the Pruisken action in the integer quantum Hall system.

\subsection{${\rm Z}_2$ index from ``half'' Wess-Zumino-Witten term}
\label{appendix_z_2 from_wess_zumino_witten}

The alternative formulation of the ${\rm Z}_2$ index borrows 
concepts from the construction of topological Wess-Zumino-Witten (WZW) actions. 
It builds on a continuous deformation of the mapping $U_\bold{k}$ to the constant matrix $\sigma_2$.
Generally, such deformation can only be found
if one relaxes the original symmetry constraint, 
and allows the latter to leave the coset space and 
belong to the larger unitary group. 
Being simply connected, the unitary group  leaves enough room for 
an interpolation, and indeed, 
for topologically nontrivial mappings 
the interpolation will violate time-reversal symmetry 
of the two-dimensional map at some fixed deformation parameter. 
We thus introduce the deformation
$g_U(k_0,\bold{k})$ with $g_U(0,\bold{k})=U_\bold{k}$ 
and $g_U(\pi,\bold{k})=\sigma_2$, where different symbols 
are used to recall that 
the latter belongs to the coset space while the former to the full group. 
We can then extend the deformation to a mapping 
from the $3$-torus to the coset ${\rm U}(2)/{\rm Sp}(2)$,   
by imposing time reversal symmetry on the three-dimensional map, 
 $g_U(k_0,\bold{k})=\sigma_2g_U^T(-k_0,-\bold{k}) \sigma_2$. 
This mapping can now be characterized by a winding number. 
The crucial observation then 
is that by construction 
(viz. time-reversal symmetry)  
 integrals over positive and negative half tori, 
$-\pi \leq k_0 \leq 0$ and $0 \geq k_0 \geq \pi $ respectively, 
 contribute equally to the winding. 
One may thus define the $\theta$ angle 
as $2\pi$ multiplied by half-winding number,
\begin{align}
\label{eq:theta}
\theta[U_\bold{k}] 
&= 
\frac{1}{12\pi} \int_{\cal M} 
\mathrm{tr}\left(
\Phi_U \wedge \Phi_U \wedge \Phi_U 
\right)\,\,{\rm mod}\,\, 2\pi,   
\end{align}
where 
$\Phi_U =g_U^{-1} dg_U$ and  
 the integral here is over the half $3$-torus 
${\cal M}=[0,\pi]\times [-\pi,\pi]^2$.
Notice however, that   
being constructed from a WZW action
it only depends on the boundary value, i.e. modulo $2\pi$ it is uniquely fixed by 
the original map $U_\bold{k}$.
 Eq.~\eqref{eq:theta} 
takes values $0$ or $\pi$ 
and we verify explicitly in Sections~\ref{appendix_z_2 from_wess_zumino_witten} 
that for the dynamical protocol realizing FM$_{1+1_{\rm syn}}$ 
it leads to the result Eq.~\eqref{eq:z2_QKR}.

We can now extend the above construction to introduce a topological action for 
the field mapping Eq.~\eqref{mapping_T_class_aII}. 
Following the above procedure we introduce a 
deformation of the 
original mapping 
to $\tau_3\otimes\openone_{2R}$, i.e. 
$g(x_0,\bold{x})$ 
with
$g(0,\bold{x})= Q(\bold{x})$ 
and
$g(1,\bold{x})=\tau_3\otimes\openone_{2R}$. 
Noting that $\pi_2({\rm O}(4R))=0$, we allow $g(x_0,\bold{x})$ to 
leave the coset space and belong to the orthogonal group, 
which makes it always possible to find a deformation.
In the final step 
we employ the time reversal constraint to 
extend the mapping from the half to the full $3$-torus (see Section~\ref{appendix_soft_mode_actions} for details). 
The resulting mapping is again characterized by a winding number, and 
it can be verified (see Section~\ref{appendix_soft_mode_actions}) 
that  integrals over positive and negative half tori, 
 contribute equally to the winding. 
This allows us to define 
half of a WZW action,
\begin{align}
\label{hwzw}
     \Gamma[g]
     &=
{1\over 24 \pi}  \int_{\cal M} \tr\left( \Phi_g \wedge \Phi_g\wedge \Phi_g \right),
 \end{align} 
 with $\Phi_g\equiv g^{-1}dg$ and integration over half the $3$-torus 
${\cal M}=[0,1]\times [-1,1]^2$. The above action is 
defined up to multiples of $2\pi$, and
the right hand side here recalls that 
 the WZW function only depends on the boundary value, i.e. is 
(modulo $2\pi$) uniquely determined by the original mapping $Q(\bold{x})$.  
Notwithstanding the multi-valuedness of individual terms, 
the combination of both contributions, Eq.~\eqref{chern_simons_action}, 
is unambiguously defined
$e^{-S_{\rm top}[Q]} = \pm 1$.
  Again  we notice that the action Eq.~\eqref{hwzw} and coupling constant 
  Eq.~\eqref{eq:theta} 
 are largely conditioned by the same (symmetry) principles, 
 and refer
to Sections~\ref{appendix_cft},~\ref{appendix_soft_mode_actions}, 
 and~\ref{appendix_derivation_aii_topological_term} 
 for additional details on the 
construction.

{\it Explicit parametrization:---}To demonstrate the topological 
nature of the protocol Eq.~\eqref{fm1+1}
we again concentrate on the mapping induced by 
$U_\bold{k}$. 
We introduce the third momentum $s$ that interpolates from 
$U (k_0=0,\bold k) = U_{\bold k}$ and $U(k_0=\pi,\bold k) = \sigma_0$, 
which is clearly trivial, and compute the winding number from $T^3$ to 
${\rm SU}(2)$ to determine whether the model at $k_0=0$ is topologically 
non-trivial. For that we compute the winding number
of $q = (k_0, \bold k) = \sigma_2 U(k_0,\bold k)$,
\begin{eqnarray}
\label{eq:theta1}
\theta = \frac{1}{12\pi}\int_0^\pi dk_0 
\int d^2\bold k \epsilon^{\mu\nu\rho} 
\tr (q^\dg\partial_\mu q q^\dg\partial_\nu q q^\dg\partial_\rho q ), \nonumber \\
\end{eqnarray}
using the following parametrization~\footnote{
	Note that for such construction $U(\pi/2 ,k) \to - i \sigma_3$, which is
	not yet time-reversal symmetric operator. However, a subsequent rotation $ e^{i \psi \sigma_3}(-i\sigma_3)$ with $\psi \in [0,\pi/2]$ brings it to the identity matrix $\sigma_0$. At this interval a spectral density in (\ref{eq:theta1}) is zero thereby giving no contribution to $\theta$--angle.}
\begin{equation}
U(k_0,\bold k) = \frac{U_\bold{k}-i\tan(k_0/2)\sigma_3}{\sqrt{1-2u_3 \tan(k_0/2) + \tan^2(k_0/2)}}.
\end{equation}
As one easily checks, the time-reversal condition, $\sigma_2 U^T(k_0,\bold k)\sigma_2 = U(-k_0, -\bold k,)$, in the extended $3d$ Brillouin zone is satisfied.
With the help of this parametrization we obtain
\begin{eqnarray}
\theta &=& \int \limits_0^{+\infty} dz \int d^2 \bold k \cos ^2 ({{k_1}}/{2}) \cos ^2 ({{k_2}}/{2}) \nonumber\\
	&\times&  \frac{\left(3 - \cos k_1 - \cos k_2 + 
	\cos k_1 \cos k_2\right) }{2 \pi  \left(1 + z^2 -  z\,\sin k_1 |\sin k_2| \right)^2} = \pi,
\end{eqnarray}
where the variable $z = \tan^2(k_0/2)$ was introduced.
To evaluate this integral one may check that after $k_1 \to - k_1$ and $z \to -z$ the former doesn't change. Thereby extending $z$-integration over the full real axis (with factor 1/2) one may complete it using residues.  The remaining integration over momenta $k_{1,2}$ eventually brings 
us the angle $\theta=\pi$.

\subsection{Derivation of AII topological term}
\label{appendix_derivation_aii_topological_term} 

We next derive the topological action Eqs.~\eqref{app_eq:S_AII_top_term} and \eqref{eq:Gamma_+}, 
starting out from the fermionic action 
$S_2$, given in Eq. (\ref{eq:S_12_AII}), 
\begin{align}
\label{eq:D_def}
S_2 
&=
\frac 12 (\chi_+^T, \chi_-^T)\, {\cal D}[Z]  \left(\begin{array}{c}
\chi_+ \\ \chi_-
\end{array}\right),
\end{align}
where we focus on the limit $\phi_\pm \to 0$, for which
\begin{align}
\label{eq:D_def2}
{\cal D}[Z] 
&= 
\left(\begin{array}{cc}
   s_1\otimes i \sigma_2  U^\dagger &  \tilde Z^T  \\ -\tilde Z  &   U s_1 \otimes i\sigma_2 
\end{array}\right).
\end{align}
Recalling the time reversal constraint, $\sigma_2 U \sigma_2 = U^T$, 
we notice that the matrix ${\cal D}$ is anti-symmetric, ${\cal D}^T = - {\cal D}$, 
and Gaussian integration over fermions therefore generates the Pfaffian
${\rm Pf}({\cal D})$. 
As we show next, the topological action 
Eqs.~\eqref{app_eq:S_AII_top_term} and \eqref{eq:Gamma_+}
accounts for the two possible 
signs of the Pfaffian, that is 
\begin{equation}
\label{eq:S_top__sign_Pf}
e^{-S_{\rm top}[Q]} 
= 
{\rm sgn}\, {\rm Pf}({\cal D}[Z]).
\end{equation}

To this end, we first focus
on the minimal coset ${\cal M}_R = {\rm O}(4R)/{\rm O}(2R)\times {\rm O}(2R)$ with 
$R=1$ and $\theta=\pi$. We show  
that for particular configurations 
of $Q$, respectively $Z$,
with nontrivial windings the sign of the Pfaffian is $-1$. 
For topologically trivial configurations, on the other hand, the 
sign of the Pfaffian is $+1$. 
Evoking continuity arguments, 
it then  follows that `${\rm sgn}\, {\rm Pf}({\cal D}[Z])$' remains fixed 
for all $Q$'s within the same topological sector independent of $R$. 
Indeed, the `${\rm sgn}$'-function is discrete but all configurations 
$Q$ within the same topological sector can be continuously 
deformed into each other. 
Before introducing the specific field configuration, we start out 
with a brief discussion of the 
geometric structure of the minimal coset space.

{\it Geometric structure of minimal coset space:---}Focusing on the minimal case 
$R=1$, the matrix $Z$ is a $2\times 2$ matrix with symmetry
 constraint $\tilde Z = - Z^\dagger = s_1 Z s_1$.
We then decompose the matrix 
into diffuson and Cooperon channels, $Z = Z_d + Z_c$, 
where 
\begin{align}
Z_d 
&= 
i w_0 s_0 + w_3 s_3 
= 
\left(\begin{array}{cc}
i w_0 + w_3 & \\ & i w_0 - w_3 
\end{array}\right)_{RA}, 
\\
Z_c 
&= i w_1 s_1 + i w_2 s_2
= 
\left(\begin{array}{cc}
& i w_1 + w_2  \\  i w_1 - w_2 & 
\end{array}\right)_{RA}.
\end{align}
Here and in the following matrices $s_i$ operate in particle-hole space, 
and $w_i$ are real numbers. 
The related $T$ fields are of the form
\begin{eqnarray}
T_{d,c} &=& \left(\begin{array}{cc}
1 & Z_{d,c} \\ -Z_{d,c}^\dagger & 1 
\end{array}\right) \equiv 1 + W_{d,c},
\end{eqnarray}
where 
\begin{eqnarray}
W_d = i w_0 \tau_1 \otimes s_0 + i w_3 \tau_2\otimes  s_3, 
\\
W_c = i w_1 \tau_1 \otimes s_1 + i w_2 \tau_1\otimes  s_2.
\end{eqnarray}

Using that generators $W$'s are mutually commuting, $[W_d, W_c] =0$, 
we next show that ${\cal M}_1 \simeq S_2 \times S_2/\mathbb{Z}_2$.
To this end, it is instructive to first recall that in the simpler class ${\rm A}$ the 
minimal coset is ${\rm SU}(2)/{\rm U}(1) \simeq S_2$. 
In this case the $Q$-matrix can be 
parametrized by a unit vector ${\bf n} = (n_1,n_2,n_3)$ or, 
equivalently, by two spherical angles $(\theta,\phi)$ as
\begin{align}
Q 
&=  {\bf n} \cdot {\bold \sigma} 
= 
\left(\begin{array}{cc}
n_3 & n_1 - i n_2 \\ n_1 + i n_2 & - n_3 
\end{array}\right)_{RA} 
\nonumber \\
&=
\tau_3 ( n_3 - i n_2 \tau_1 + i n_1 \tau_2 )
\nonumber \\
&\equiv 
\tau_3 \, e^{- \frac i2 \phi \tau_3} e^{ i \theta \tau_2}  e^{\frac i2 \phi \tau_3}.
\end{align}
Alternatively, one can use a rational parametrization 
which in the minimal case ($R=1$) uses a single complex variable 
$w= w_1  + i w_2$ (the identification is achieved by setting $\tilde Z = w$),
\begin{equation}
Q= \frac{1}{1+|w|^2}\left(\begin{array}{cc}
1-|w|^2 & 2w^* \\ 2w  & -(1-|w|^2) 
\end{array}\right)_{\rm RA},
\end{equation}
and defines the stereographic projection ${\bf n}(w)$ from $\mathbb{C} \to S_2$,
\begin{equation}
n_1 + i n_2 = \frac{2 w}{1+|w|^2}, \qquad n_3 = \frac{1-|w|^2}{1+|w|^2}.
\end{equation}

With this class ${\rm A}$ example in mind, we can  use the same mappings 
to define diffusion and Cooperon spheres. 
To this end, we define for the diffuson the matrices
\begin{align}
&\Gamma_1^d = \tau_1\otimes  s_0, 
\quad 
\Gamma_2^d = - \tau_2\otimes  s_3,
\quad
\Gamma_3^d = - \tau_3\otimes  s_3, 
\end{align}
satisfying $\Gamma_i^d \Gamma_j^d 
= i \epsilon_{ijk} \Gamma_k^d$,  
and  $W_d = i w_0 \Gamma_1^d - i w_3 \Gamma_2^d$, 
where $\Gamma_3^d$ was introduced to complete the underlying 
${\rm SU}(2)$ algebra. 
Defining then the diffuson sphere $Q_d = T_d \tau_3 T_d^{-1}$, it 
can be verified that 
\begin{align}
Q_d 
&= T_d \tau_3 T_d^{-1} = \tau_3( n_3^d- i n_2^d \Gamma_1^d + i n_1^d \Gamma_2^d ) 
\nonumber\\
&= 
\tau_3 \, e^{- \frac i2 \phi_d \Gamma_3^d} e^{ i \theta_d \Gamma_2^d}  e^{\frac i2 \phi_d \Gamma_3^d},
\end{align}
where the unit vector ${\bf n}_d = (n_1^d,n_2^d,n_3^d)$ 
and angles $(\theta_d, \phi_d)$   
follow from the stereographic projection 
\begin{equation}
n_1^d + i n_2^d 
= 
\frac{2 w_d}{1+|w_d|^2}, 
\quad
n_3^d 
= 
\frac{1-|w_d|^2}{1+|w_d|^2}, 
\end{equation}
with $w_d = w_3 + i w_0$.

For the Cooperon sphere we introduce a different 
set of ${\rm SU}(2)$ matrices mutually commuting with those of the 
diffuson,
\begin{align}
&\Gamma_1^c = \tau_1\otimes  s_2, 
\quad
\Gamma_2^c = -  \tau_1\otimes  s_1, 
\quad
\Gamma_3^c = - \tau_0\otimes  s_3, 
\end{align}
satisfying
$\Gamma_i^c \Gamma_j^c = i \epsilon_{ijk} \Gamma_k^c$ and 
$[\Gamma_i^d, \Gamma_j^c] = 0$, 
and $W_c = i w_2 \Gamma_1^c - i w_1 \Gamma_2^c$. 
Introducing then the Cooperon sphere
$Q_c = T_c \tau_3 T_c^{-1}$, 
we notice that
\begin{align}
Q_c 
&= T_c \tau_3 T_c^{-1} = \sigma_3( n_3^c- i n_2^c \Gamma_1^c + i n_1^c \Gamma_2^c ) 
\nonumber\\
&= 
\tau_3 \, e^{- \frac i2 \phi_c \Gamma_3^c} e^{ i \theta_c \Gamma_2^c}  e^{\frac i2 \phi_c \Gamma_3^c},
\end{align}
and the unit vector ${\bf n}_c = (n_1^c,n_2^c,n_3^c)$ and angles 
$(\theta_c, \phi_c)$ 
can again be obtained via 
stereographic projection
\begin{equation}
n_1^c + i n_2^c = \frac{2 w_c}{1+|w_c|^2}, 
\,\, 
n_3^c = \frac{1-|w_c|^2}{1+|w_c|^2}, 
\end{equation}
with $w_c = w_1 + i w_2$. 

Finally, we can express the full $Q$-matrix as
\begin{equation}
Q({\bf n}_d, {\bf n}_c) = T_c T_d \tau_3 T_d^{-1} T_c^{-1}  = Q_c({\bf n}_c) \tau_3 Q_d({\bf n}_d),  
\end{equation}
and notice that the odd parity properties $Q_{d,c}(-{\bf n}_{d,c}) = - Q_{d,c}({\bf n}_{d,c})$ 
leave a sign-ambiguity $Q(-{\bf n}_d, -{\bf n}_c) = Q({\bf n}_d, {\bf n}_c)$. 
We thus arrive at the stated isomorphism 
${\cal M}_1 \simeq S_2 \times S_2/\mathbb{Z}_2$.

{\it Evaluation of the Pfaffian:---}We next specify 
$Q = Q_d({\bf n}_d)$ and $Q = Q_c({\bf n}_c)$  
to non-trivial mappings from the torus $T^2$ onto 
diffuson and Cooperon spheres. That is, we choose  
configurations with finite windings $\Gamma[Q_{d,c}]=\pi$,
and show that in these cases 
the sign of the Pfaffian~(\ref{eq:S_top__sign_Pf}) is $-1$.
More specifically, we consider 
Pruisken's instantons with windings $W=1$ and $-1$.
In stereographic coordinates 
$w=x_1 + i x_2$, respectively, $w=(x_1 + i x_2)^{-1}$, and  
we  show that the WZW action for the configurations 
indeed give $\pi$. 
To this end, we  extend  $Q_{d/c}$ to the group element $g(x_0, {\bf x}) \in {\rm O}(4)$ defined in three dimensions, such that
$g(0,{\bf x})=Q_{d/c}({\bf x})$ and $g(\pi/2,{\bf x})=\tau_3$. 
Such extension can be achieved in two steps. 
First we introduce
\begin{eqnarray}
g(x_0, {\bf x}) &=&  Q_{d/c}({\bf x}) \cos x_0 + i \tau_3 \Gamma_3^{d/c} \sin x_0, \\
g^{-1}(x_0,{\bf x}) &=& Q_{d/c}({\bf x}) \cos x_0 - i \tau_3 \Gamma_3^{d/c} \sin x_0,
\end{eqnarray}
and notice that all matrices $\Gamma_i^{d/c}$ satisfy the symmetry condition  
\begin{equation}
\label{eq:Gamma_bar}
\bar\Gamma_i^{d/c} = C^T (\Gamma_i^{d/c})^T C = -\Gamma_i^{d/c}, \qquad C= \tau_3 \otimes s_1,
\end{equation}
implying that $\bar g = g^{-1}$, i.e. $g \in {\rm O}(4)$ as required.
In the above parametrization, the final value $g(\pi/2, {\bf x})=i \tau_3 \Gamma_3^{d/c}$ 
is still different from $\tau_3$, and in the second step 
we introduce an additional rotation 
$e^{ -i \psi \Gamma_3^{d/c}} g(\pi/2, {\bf x})$ with $\psi\in [0,\pi/2]$
which brings the latter to $\tau_3$. 
Notice, however, that this rotation is inessential 
for the evaluation of the WZW action. 
Using that we have a one-to-one mapping from the 
$2d$ space (or torus $T^2$) onto the sphere $S_2$
we can evaluate the WZW action as
\begin{align}
\label{wzwg}
\Gamma[g] 
&= 
\frac{1}{8\pi} \int\limits_0^{\pi/2} dx_0 \int d^2 \bold x \, {\rm tr} (g^{-1} \partial_{x_0} g \,[g^{-1} \partial_{x_1} g, g^{-1} \partial_{x_2} g]) 
\nonumber\\
&= 
\frac{1}{8\pi} \int\limits_0^{\pi/2} dx_0 \int\limits_0^{\pi} d\theta 
\int\limits_0^{2\pi} d\phi \, {\rm tr} (g^{-1} \partial_{x_0} g \,[g^{-1} \partial_{\theta} g, g^{-1} \partial_{\phi} g])
\nonumber\\
&= \frac{1}{\pi} \int\limits_0^{\pi/2} dx_0 \int\limits_0^{\pi} d\theta 
\int\limits_0^{2\pi} d\phi\, (\cos x_0)^2 \sin\theta = \pi, 
\end{align}
which shows that the considered mappings 
$Q_{d/c}({\bf x})$ belong to a non-trivial homotopy class.

Finally, we show that both mappings give negative signs for the Pfaffian.
To evaluate ${\rm Pf}({\cal D}[Z])$ on the configuration 
$Q_d({\bf x})$, defined by a unit vector
${\bf n}_d$ with non-trivial winding $W_d = \pm 1$, 
we start out from the explicit expression
of ${\cal D}$ in Eq.~(\ref{eq:D_def2}) 
\begin{equation}
{\cal D} =\left( \begin{array}{cccc}
0 & i \sigma_2 U^\dagger & i w_0 - w_3 & 0\\
i \sigma_2 U^\dagger & 0 & 0 & i w_0 + w_3 \\
-i w_0 + w_3 & 0 & 0 &  i U \sigma_2 \\
0 & - i w_0 - w_3 & i U \sigma_2 & 0 
\end{array} \right).
\end{equation}
We then subject ${\cal D}$ 
to an orthogonal transformation defined by 
 the matrix 
\begin{equation}
B =\left( \begin{array}{cccc}
1 & 0 & 0 & 0\\
0 & 0 & 0 & 1 \\
0 & 0 & 1 & 0 \\
0 & 1 & 0 & 0 
\end{array} \right),
\end{equation}
which swaps the 2nd and the 4th rows/columns.
This brings ${\cal D}$ into block off-diagonal form,
\begin{align}
\tilde {\cal D} 
&= 
B {\cal D} B^T = 
\left( \begin{array}{cc}
0 & A \\
- A^T &  0
\end{array} \right), 
\\
A 
&= 
\left( \begin{array}{cc}
i w_0 - w_3 &  i \sigma_2 U^\dagger \\
i U \sigma_2 & - i w_0 - w_3  
\end{array} \right).
\end{align}
At this stage we use 
that 
\begin{equation}
{\rm Pf}(\tilde {\cal D}) = \left(\prod_{{\bf k}}{\rm det}\, B \right) 
{\rm Pf}({\cal D}) = (-1)^{m(m-1)/2}\, {\rm det} A,
\end{equation}
where $m$ is the size of the matrix $A$. 
In the given case $m=4M$ with $M$ some integer, 
and the sign reduces to $+1$. 
Similarly, each generic momentum ${\bf k}$ has its TRS image $-{\bf k}$, 
and the number of time-reversal invariant momenta is 4, implying  
that a negative determinant ${\rm det}\, B = -1$ plays no role. 
Finally, upon introducing the complex field 
$w_d({\bf x}) = w_3({\bf x}) + i w_0({\bf x})$, 
the evaluation of ${\rm det} A $  
can be reduced to 
\begin{align}
\label{eq:Det_A}
{\rm det} A 
&= {\rm det} \left( \begin{array}{cc}
- w_d^\dagger &  i \sigma_2 U^\dagger \\
i U \sigma_2 & - w_d  
\end{array} \right) = {\rm det}  \left( \begin{array}{cc}
i U \sigma_2 & -w_d \\
- w_d^\dagger & i \sigma_2 U^\dagger  
\end{array} \right) 
\nonumber\\
&= 
{\rm det} ( 1 + U^\dagger w_d U w_d^\dagger). 
\end{align}
This is a familiar form, we already encountered previously in the derivation of a class A action, see Ref.~\cite{kim2020quantum}. 
In particular, the sign of this determinant 
can be defined via the (imaginary) Pruisken action,
\begin{align}
\label{eq:detA_diff_pi}
{\rm sgn} ( {\rm det} A) 
&= 
\exp\left\{ \frac{\theta}{16\pi} \int d^2 \bold x \, 
\epsilon^{ij}{\rm tr}( Q_d \partial_i Q_d \partial_j Q_d) \right\}  
\nonumber\\ 
&=
e^{ i \theta W_d} = e^{i\theta} = -1,
\end{align}
where the very last relation holds if $U= U_{\bold k}$ is topologically 
non-trivial with $\theta = \Gamma[\hat U_{\bold k}] = \pi$.  This validates the 
representation of the topological term 
as the product of two WZW terms in momentum and real space: when the configuration of field $Q_d$ is nontrivial, the Pruisken term counts its winding in $2d$ space, and as shown in Eq.\eqref{wzwg}, the WZW term is identically nontrivial. 
Provided that the Floquet operator has finite winding, 
and any nontrivial content of a field configuration 
$Q = Q_c(\bold n_c)\tau_3 Q_d(\bold n_d)$, 
can be smoothly deformed into the diffuson or Cooperon sector, $Q_d(\bold n_d)$ respectively $Q_c(\bold n_c)$,  
the topological term coincides with the sign-factor 
`$\rm{sgn} Pf(D[Z])$', as we set out to show. 

We conclude this section by completing the Cooperon part of the derivation. 

The evaluation of the Pfaffian for a $Q$-matrix with non-trivial 
twist in the Cooperon channel proceeds along the same lines. 
In this case the antisymmetric matrix ${\cal D}$ simplifies to
\begin{equation}
{\cal D} =\left( \begin{array}{cccc}
0 & i \sigma_2 U^\dagger & 0 & i w_1 - w_2\\
i \sigma_2 U^\dagger & 0 & i w_1 + w_2 & 0 \\
0 & -i w_1 - w_2 & 0 &  i U \sigma_2 \\
-i w_1 + w_2 & 0 & i U \sigma_2 & 0 
\end{array} \right),
\end{equation}
and the orthogonal transformation that exchanges 
2nd with 3rd columns/rows brings 
it to the block off-diagonal form,
\begin{eqnarray}
\tilde {\cal D} &=&  
\left( \begin{array}{cc}
0 & A \\
- A^T &  0
\end{array} \right), \nonumber\\
A &=& \left( \begin{array}{cc}
 i \sigma_2 U^\dagger & i w_1 - w_2 \\
  - i w_1 - w_2 & i U \sigma_2  
\end{array} \right).
\end{eqnarray}
With a spatially dependent complex field 
$w_c({\bf x}) = w_1({\bf x}) + i w_2({\bf x})$ the Pfaffian becomes
\begin{align}
{\rm Pf}(\tilde {\cal D}) 
&= {\rm det} A = {\rm det}\left( \begin{array}{cc}
i \sigma_2 U^\dagger & i w_c \\
- i w_c^\dagger & i U \sigma_2  
\end{array} \right) 
\nonumber\\
&= 
{\rm det} ( 1 + U w_c U^\dagger w_c^\dagger),
\end{align}
which,
similar to the diffuson channel,
 gives a negative sign for the Pfaffian 
\begin{align}
{\rm sgn} ( {\rm det} A) 
&= 
\exp\left\{ \frac{\theta}{16\pi} \int d^2 \bold x \, 
\epsilon^{ij}{\rm tr}( Q_c \partial_i Q_c \partial_j Q_c) \right\} 
\nonumber\\
&=
e^{ i \theta W_c} = e^{i\theta} = -1,
\end{align}
provided topological angle $\theta=\pi$, cf. Eq.~(\ref{eq:detA_diff_pi}).

%
\end{document}